\newcommand{\mat}[1]{\begin{pmatrix}#1\end{pmatrix}}

\documentclass[
 reprint,
 amsmath,amssymb,
 aip,
 superscriptaddress
]{revtex4-1}

\usepackage{graphicx}
\usepackage{dcolumn}
\usepackage{bm}
\usepackage{float}
\usepackage{hyperref}
\usepackage{csquotes}
\usepackage[dvipsnames]{xcolor}
\usepackage{gensymb}
\usepackage{pbox}
\usepackage[all]{nowidow}
\usepackage{textcomp}
\usepackage{mathrsfs}

\usepackage{color}

\begin{document}

\preprint{The Journal of Chemical Physics}

\title{Matching Crystal Structures Atom-to-Atom}

\author{F\'elix Therrien}
\affiliation{Colorado School of Mines}
\affiliation{National Renewable Energy Laboratory}
\author{Peter Graf}
\affiliation{National Renewable Energy Laboratory}
\author{Vladan Stevanovi\'c}
\email{vstevano@mines.edu}
\affiliation{Colorado School of Mines}
\affiliation{National Renewable Energy Laboratory}
\date{\today}

\begin{abstract}
Finding an optimal match between two different crystal structures underpins many important materials science problems, including describing solid-solid phase transitions, developing models for interface and grain boundary structures. In this work, we formulate the matching of crystals as an optimization problem where the goal is to find the alignment and the atom-to-atom map that minimize a given cost function such as the Euclidean distance between the atoms. We construct an algorithm that directly solves this problem for large finite portions of the crystals and retrieves the periodicity of the match subsequently. We demonstrate its capacity to describe transformation pathways between known polymorphs and to reproduce experimentally realized structures of semi-coherent interfaces. Additionally, from our findings we define a rigorous metric for measuring distances between crystal structures that can be used to properly quantify their geometric (Euclidean) closeness.
\begin{description}
\item[Keywords]
Structure matching, Polymorphism, Phase transitions, Heterojunctions
\end{description}
\end{abstract}

\maketitle

\section{Introduction}
%
Establishing an optimal match between two different crystal structures with respect to some cost function is a problem that cuts across the entire field of materials science. Perhaps the most evident example is the process of finding a suitable substrate to epitaxially grow a material \cite{brune:2014, rondo_NC:2016, persson_ACS:2016}. Similarly, when studying interfaces between different phases (heterojunctions) one might be interested in the alignment and the bonding pattern between the two phases. Another important example lies in finding minimal energy pathways between different polymorphs. The initial and final structures are known, but the transformation from one to the other is not. To even begin to describe it, one needs to find the best way to map every atom of the initial structure to its counterpart in the final structure and to optimally align the structures. Once the mapping and alignment are established other methods such as the Solid State Nudge Elastic Band \cite{henkelman_2000, sheppard_2012, caspersen_2005, qian_2013} can be used to determine the energetics of the transition.

In regard to interfaces, many have worked on methods to find and characterize the coincidence of lattices and orientation relationships between phases and grains. Several different approaches were developed e.g., the O-lattice theory part of the CSL/DSC Lattice Model \footnote{CSL: Coincidental Site Lattice, DSC: \underline{D}isplacement of one crystal Lattice with respect to the second causes a pattern \underline{S}hift which is \underline{C}omplete} 
\cite{bollmann1, bollmann_1974, smith_pond_1976, balluffi_brokman_king_1982}, the Edge to Edge Model \cite{zhang_kelly_1998, zhang_kelly_easton_taylor_2005}, the Coincidence of Reciprocal Lattice Points (CRLP) model \cite{ikuhara_pirouz_1996}, methods based on the Zur Algorithm \cite{zur1984, persson_ACS:2016, mathew_2016} and the work of Jelver et al. \cite{jelver2017determination}. While the O-lattice theory suffers from a lack of predictive capabilities, the other approaches do have the ability to predict orientation relationships, but they do not match the full structures. The Edge to Edge model only considers high density (nearly close packed) planes and directions whereas the CRLP and Zur Algorithm only match the underlying lattices  of the structures, and not the atoms inside them. Jelver et al. presented a crystal matching method that maps atoms inside a combination of the unit cells of the two structures which, as explained further in this section, has some important limitations.

Matching is also closely related to measuring distances between crystal structures. Indeed, any definition of a distance metric requires establishing some correspondence between their atoms. For finite systems such as molecules, 
Sadeghi and Goedecker \cite{sadeghi_2013} defined a distance as the minimal $l_2$-norm of the vector joining the molecules in the configuration space of atoms with respect to both their relative positions (alignment) and the permutation of atomic indices \footnote{The $l_2$-norm in configuration space equivalent to the Forbenius norm of the position matrix}. However, the configuration space of periodic systems is, strictly speaking, ill-defined because of the infinite number of dimensions. Moreover, the permutation degeneracy in labeling atoms also poses problems. Both of these make the matching of crystal structures challenging and the definition of the distance metric between periodic structures elusive.

In structure predictions, for the purpose of identifying similar (close) structures Oganov elegantly circumvented this problem by introducing the so-called fingerprint function constructed to reflect the short-range order (coordination in various shells, etc.) and defining the distance metric between crystal structures with respect to it \cite{oganov_JCP:2009}. Various other fingerprint functions and representations have been proposed since \cite{bartoK_2013, yang_2014, de_2016, zhu_2016}. Although they can efficiently measure the similarity between structures, those methods do not establish a one-to-one correspondence between each atom of the structures, nor do they provide the optimal alignment between them. Therefore, the question whether a Euclidean distance metric in the configuration space of atoms ($l_2$-norm) and its corresponding matching can be defined for periodic systems remains open.
\begin{figure}
\includegraphics[width=1.0\linewidth]{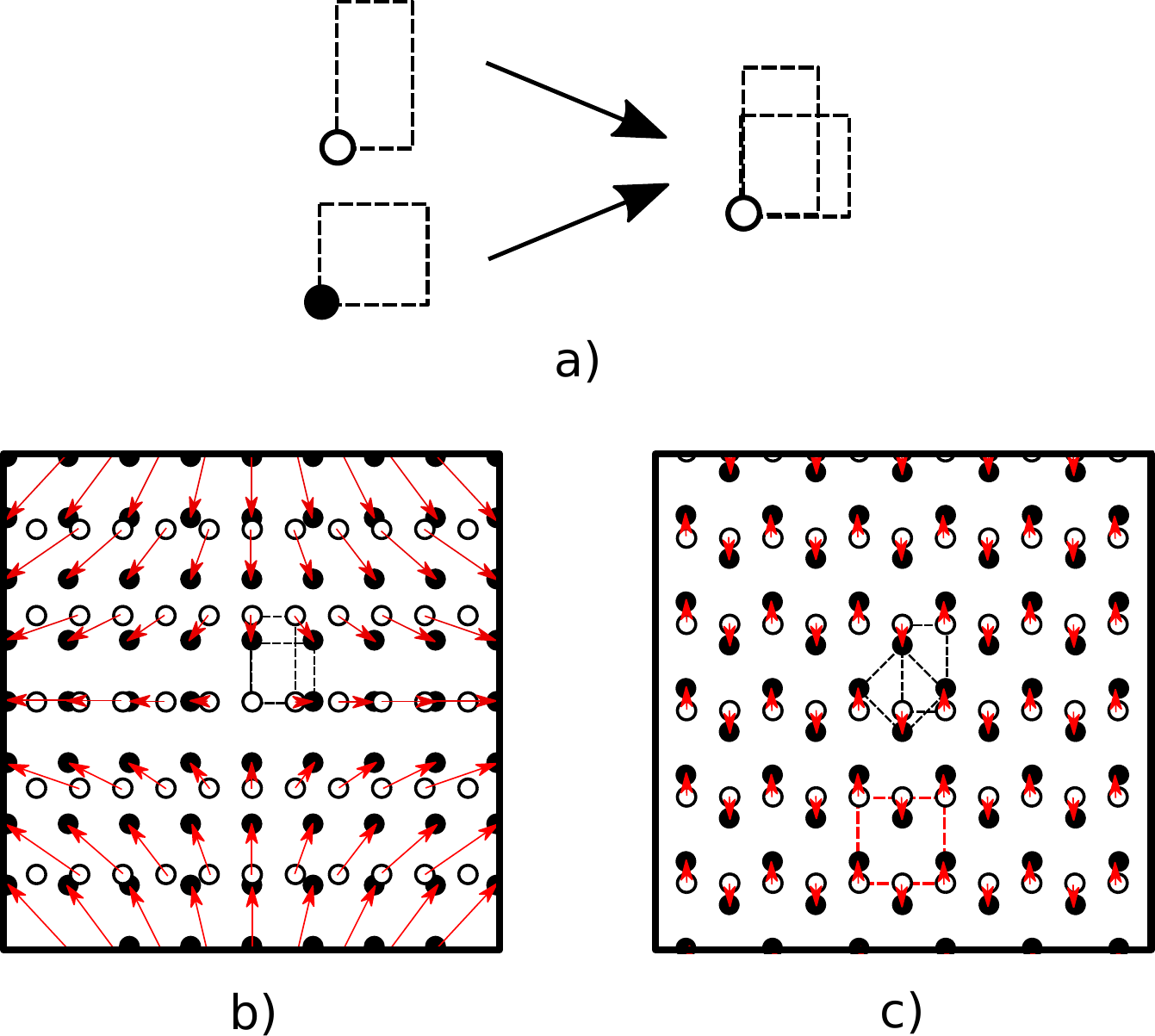}
\caption{\label{fig:limitations}
Drawbacks of a method that relies on matching some choice of the unit cells of the two crystals. (a) Visual representation of the matching inside one cell for a simple 2D example where the cells have the same area. (b) The two overlaid crystal structures using the same matching. (c) The same crystal structures overlaid such that the distance (red arrows) does not increase away from the center; the red dashed rectangle shows the scale (periodic unit) of the match.}
\end{figure}

An intuitive way to go about this problem is to take advantage of the periodicity by matching and measuring the distance between atoms inside the unit cells of the two structures. This inevitably leads to the obstacles illustrated in Fig.~\ref{fig:limitations}. Panel (a) shows two 2D crystals with different unit cells each having one atom per cell. If, for example, one wishes to minimize the distance between the atoms, a naive way to match these two structures would be to align their unit cells so that the atoms overlap. This would produce a distance of zero. Other atoms would simply be mapped based on the correspondence between the unit cells. However, this mapping would lead to the distances between the corresponding atoms diverging as one moves away from the two perfectly aligned unit cells at the center of Fig.~\ref{fig:limitations}(b). There is, however, a solution to this particular problem shown in Fig.~\ref{fig:limitations}(c) that does not suffer from this divergence and that produces equal and finite distances between all corresponding atoms. This solution yields a much shorter total distance if large portions (grains) of the two crystals are considered. While matching larger supercells and taking advantage of their Niggli-Santora-Gruber reduced cell \cite{niggli1928handbuch, santoro1970determination, krivy1976} might seem like a solution to this particular example, choosing the size of the supercells and comparing distances between different sizes remains an issue.

One way to robustly find solutions such as the one from Fig.~\ref{fig:limitations}(c) is by disregarding the periodicity in the two structures. If it exists, the periodicity of the mapping itself can be retrieved subsequently. For this reason, as we will explain, our algorithm is constructed to use large sections of the two crystals and to minimize the total distance traveled by \textit{all} the atoms. Consequently, the choice of a unit cell has no impact on the final result and the periodic unit of the transformation emerges naturally. Any other method to match two crystal structures that relies on matching some choice of the unit cell, including our previous work \cite{stevanovic2018} as well as work of others \cite{lonie_2012, capillas_2007, jelver2017determination, larsen2017structural}, will suffer from the problem illustrated in Fig.~\ref{fig:limitations}.

With the aforementioned considerations, we formulate the problem of matching crystal structures in the following way: given the positions of all the atoms in the two crystals, $\{\vec{a}_i | i=1,\dots,N \}$ and $\{\vec{b}_j | i=1,\dots,N\}$, what is the best atom-to-atom mapping $p_{min}$ (permutation of atomic indices) and the best alignment of the two structures (linear transformation $Q_{min}$ and translation $\vec{t}_{min}$) that minimize a given distance (cost) function $d$? This is equivalent to solving the following equation when $N \to \infty$: 
\begin{equation}\label{eq:min}
    p_{min}, Q_{min},\vec{t}_{min} = \underset{p, Q, \vec{t}}{\text{argmin}} \sum_{i}^{N} d(\,\, \vec{a}_{i} \,,\, Q\vec{b}_{p(i)} + \vec{t} \,\,).
\end{equation}
Formulating the problem in this way has the advantage of making the solution method easily adaptable to any given cost function, being the sum of Euclidean distances ($l_2$-norms) between the atoms or some other function depending on the particular problem or application. Posed as such, matching the structures is equivalent to doing a Point Set Registration (PSR)\cite{zhu2019review}, a well-studied process used in computer vision and pattern recognition. Our algorithm is inspired by PSR methods.

Herein, we describe our structure matching algorithm in detail and showcase its applications to phase transformations and semi-coherent interfaces. We demonstrate that it robustly reproduces known results for several well-studied polymorphic transformations. It also seamlessly reproduces and explains the experimentally observed semi-coherency and orientation relationships for the interfaces between Si and SiC, and Ni and Yttria Stabilized Zirconia (YSZ). Finally, drawing upon the results of our crystal matching algorithm we discuss and propose a proper Euclidean metric between infinitely periodic crystal structures.
%
\section{The Algorithm}
%
We start by introducing nomenclature used in this paper. Crystal structures are represented by three objects: the unit cell matrix $(C)$, the $3\times N$ matrix of atomic positions $(P)$ where $N$ is the number of atoms in the unit cell, and the $1\times N$ list ($L$) of the symbols of chemical elements occupying those positions. It is understood that the order of elements in the matrix $P$ and the list $L$ is the same. Finally, we combine these three into a single crystal structure object:
\begin{equation*}
    A \equiv \{ C_A , P_A, L_A\}.
\end{equation*}
It is important to note that there exist an infinite number of representations of the same structure due to the arbitrariness in the choice of the periodic unit (unit cell).
\begin{figure}
\includegraphics[width=1.0\linewidth]{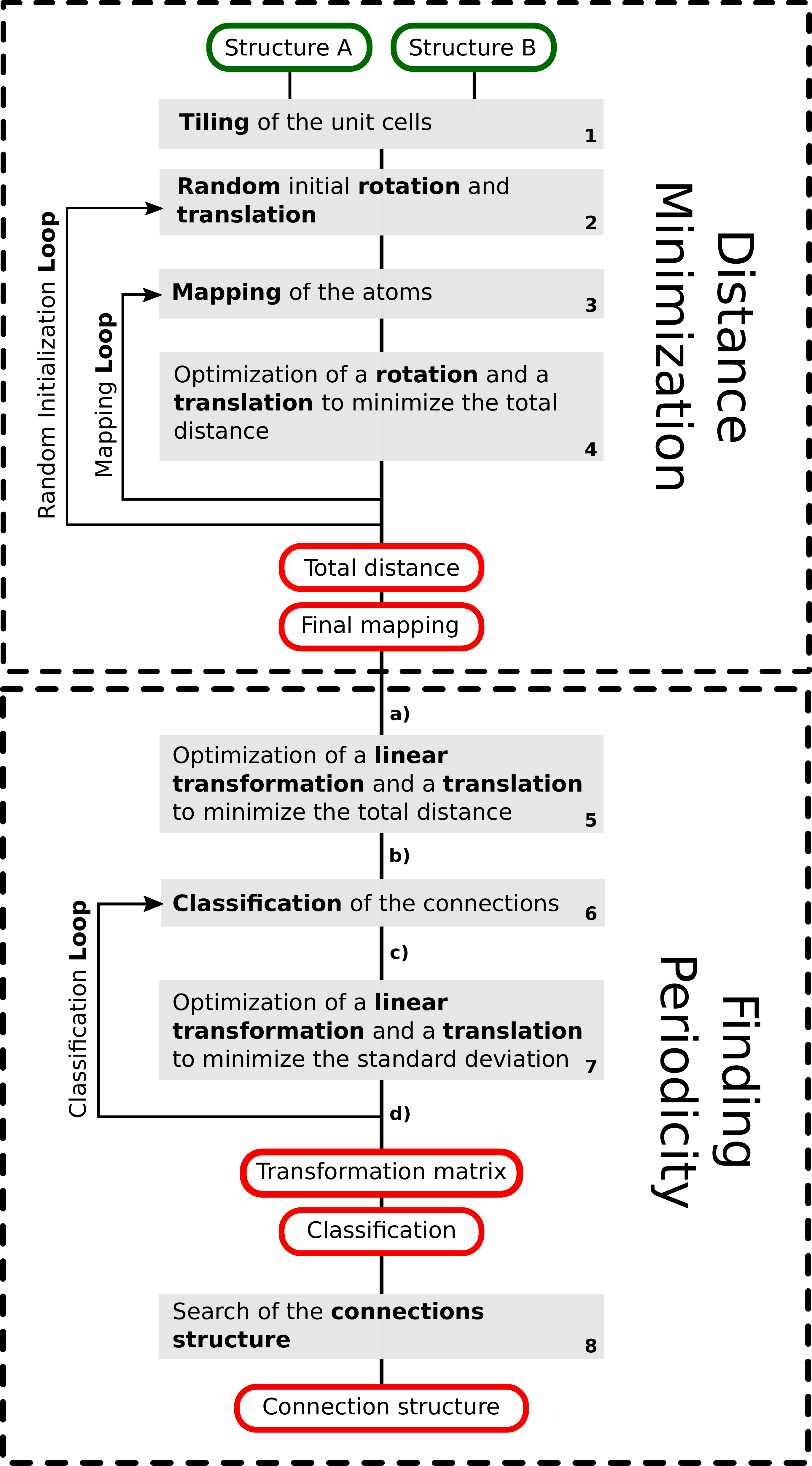}
\caption{\label{fig:algorithm}
Workflow of the algorithm. Green bubbles represent inputs, and red bubbles represent outputs. The letters correspond to the panels of Fig.~\ref{fig:post-processing}}
\end{figure}
%
\subsection{Distance Minimization}
%
The first step to solving equation \eqref{eq:min} in practice is to make the two structures we would like to match finite. The structures $A$ and $B$ are the primary input to our algorithm along with the size of the finite sections of the structures and the distance function $(d)$ to optimize. We make $A$ and $B$ finite by cutting out spherical sections around the origins of the two structures ensuring that the stoichiometry is preserved in each structure and that they both have the same number of atoms. Making the sections approximately spherical is done by selecting the atoms that are the closest to a central point in a process called \textit{tiling} as denoted in the flowchart of our algorithm shown in Fig.~\ref{fig:algorithm}.

Next, the structures are brought to the same geometric centers and an initial random rotation $Q_0$ and a random translation $\vec{t}_0$ are applied to one of them (step 2, see Fig.~\ref{fig:algorithm}). The structure that is rotated and translated is labeled ``mapping structure'' ($B$), to which all subsequent geometric transformations will be applied and the other one is labeled ``mapped structure'' ($A$) and it remains fixed in space. The initial translation $\vec{t}_0$ is constrained to within one unit cell of the mapped structure~($A$).

After this initial alignment, atoms in the two structures are mapped to each other, i.e., the permutation $p_1$ of atom indices in the mapping structure is chosen such that the distance function is minimized (step 3). The details of the mapping procedure are provided in section~\ref{mapping}. Next, for this particular atom-to-atom map, the distance between the two structures is minimized with respect to rotations $Q_1$ and translations $\vec{t}_1$ using a gradient descent (step 4). It is important to note that $Q$ does not need to be a rigid rotation. Depending on the application, it can also contain a certain amount of deformation such that $1-\varepsilon < det(Q) < 1+\varepsilon$. At this point, the atom-to-atom map $p_1$ is not necessarily optimal, since it was established before the translation and rotation were optimized. Step 3 is therefore repeated using the new alignment $Q_1$ and $\vec{t_1}$ to obtain $p_2$. Then, with this new mapping, the algorithm finds $Q_2$ and $\vec{t}_2$ (step 4), remaps again and so on, iteratively, until the $p,Q$ and $\vec{t}$ stop changing. This iterative procedure can be mathematically formulated as: 
\begin{equation}
\begin{split} \label{eq:iter_1}
    p_j  =  \underset{p}{\text{argmin}} \sum_{i}^{N} d\left(\vec{a}_i,Q_{j-1}\vec{b}_{p(i)} + \vec{t}_{j-1}\right),  \\
    \vec{t}_{j}, Q_{j} =  \underset{\vec{t}, Q}{\text{argmin}} \sum_{i}^{N} d\left(\vec{a}_i, Q\vec{b}_{p_j(i)} + \vec{t} \, \right),
\end{split}
\end{equation}
where the index $j$ is the iteration number of the Mapping Loop from Fig.~\ref{fig:algorithm}. At the end of the Mapping Loop the algorithm has reached a local minimum. In order to find the global minimum, one needs to explore the dependence of the results on the random initialization. This we do by constructing an outer Random Minimization Loop by repeating the whole procedure (i.e., steps 2,3 and 4) a large number of times until the best local minimum stops changing.

The distance minimization part of the algorithm is a form of iterative closest point (ICP)\cite{besl_1992}, a classic scheme to solve point set registration (PSR) problems that iteratively associates two sets of points and minimizes the distance between them. In our case, the data association is done by solving the assignment problem using the Khun-Munkres algorithm (see section~\ref{mapping}). To the best of our knowledge, only a few ICP algorithms have used this method to associate data \cite{pan2018iterative, bhandarkar2004surface}. Moreover, our problem differs widely from most PSR problems for the following reasons: (1) The crystals are by definition featureless since they are infinite periodic sets of points. Macroscopic features cannot be used to reduce the number of points or to find an approximate result like it is done in many PSR algorithms including refs \cite{pan2018iterative, bhandarkar2004surface} (2) The two structures are three-dimensional (in the case of phase transformations), i.e., points occupy the full 3D space, they do not represent the surface of a 3D object like in most PSR problems \cite{pan2018iterative, bhandarkar2004surface, maiseli2017recent} (3) The two structures are in many cases inherently different; they are not supposed to be similar whereas typical PSR problems aim to match different representations of a same object e.g., modeled and measured data, data from different instruments, two pieces of a broken object, etc. \cite{pan2018iterative, bhandarkar2004surface, maiseli2017recent} (4) Every point, its mapping and the distance function to minimize are physically significant. For that reason, no point can be ignored or considered as an outlier, each atom needs to map to exactly one atom (in the case of phase transformations) and the cost function must be relevant to the problem at hand (phase transformation, interfaces, similarity measurement, etc.).
Therefore, even though the distance minimization part of our algorithm is a form of ICP, other existing PSR algorithm could not be used \textit{directly} to find the optimal one-to-one mapping and alignment between crystal structures.
%
\subsection{Atom-to-Atom Mapping}\label{mapping}
%
\begin{figure}
\centering
\includegraphics[width=0.9\linewidth]{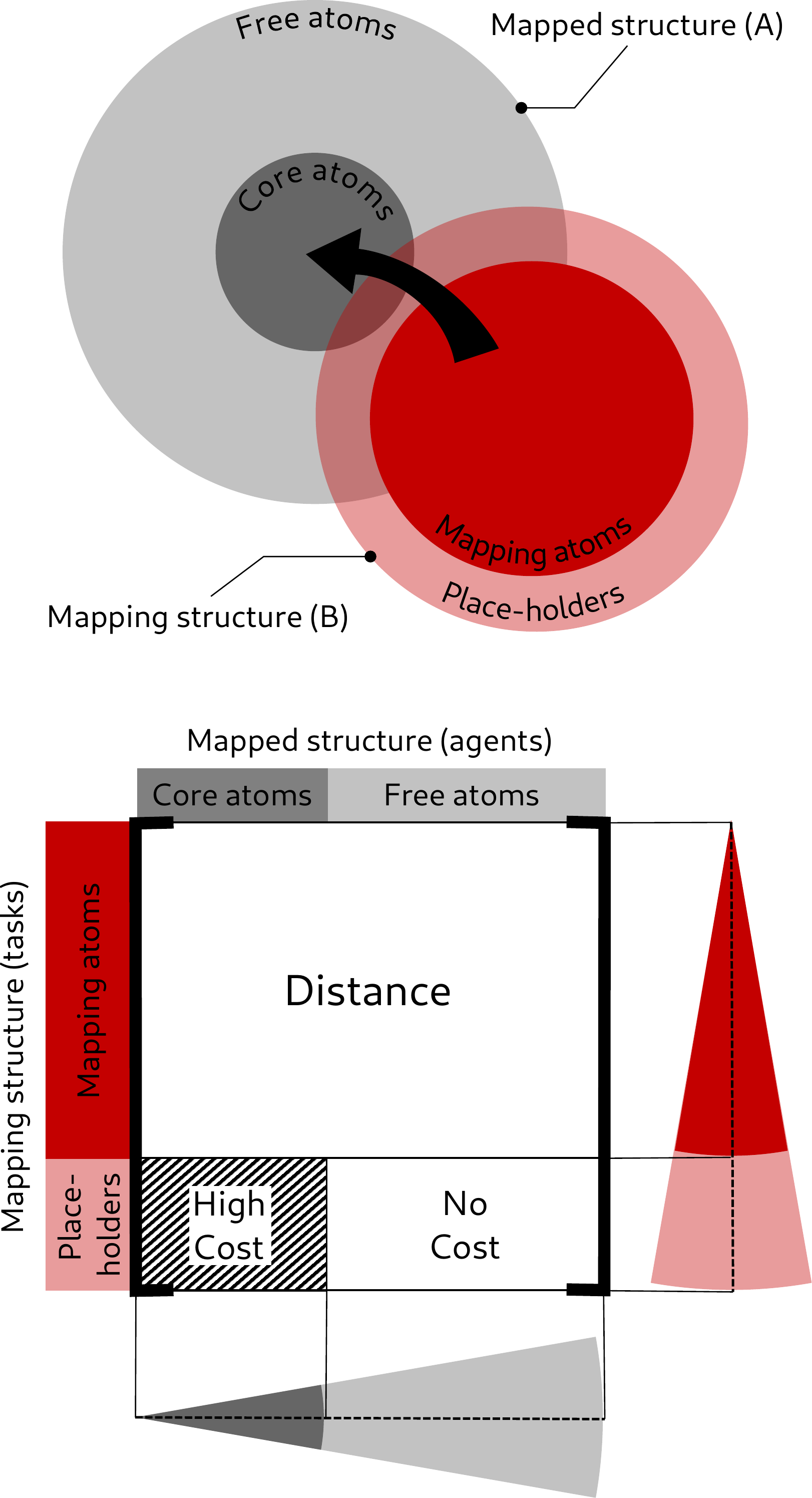}
\caption{\label{fig:cost_matrix}
Mapping of the structures. The top part shows qualitatively how the structures are mapped. The bottom part depicts the cost matrix of the assignment problem. Regions of different colors in the structures correspond to different costs in the matrix (see text for details).}
\end{figure}
Let us explain in more details how the permutation $p$ of atomic indices is optimized. At a given Mapping Loop iteration, the position of each atom in both sections of the $A$ and $B$ structures is known. The goal is to assign each atom of the mapping structure to an atom in the mapped structure such that the sum of the distances between the pairs of corresponding atoms is minimized. As cleverly noted by Sadeghi and Goedecker \cite{sadeghi_2013}, this is exactly analogous to the assignment problem, a well-studied mathematical problem for which there exist an exact solution that can be computed in polynomial time \cite{kuhn1955}.

The assignment problem consists of finding an optimal way to assign agents to tasks, e.g., clients (tasks) to their taxis (agents) such that the total distance traveled by all the taxis (cost) is minimized. Once the position of each atom is known, the structure mapping problem is exactly equivalent to the assignment problem. The algorithm needs to assign each atom of the mapping structure (or task), shown in red in Fig.~\ref{fig:cost_matrix}, to an atom in the mapped structure (or agent), shown in gray, such that the total distance (cost) is minimized. The naive route to solving this problem is to try all possible assignments of atoms, but this operation scales as $N!$. Instead, our algorithm uses the existing Kuhn-Munkres method \cite{kuhn1955} (also known as the Hungarian Algorithm) that solves the assignment problem in polynomial time. This algorithm takes as an input a cost matrix which consists of distances between each possible pair of atoms between the two structures.
\begin{figure*}[t!]
\includegraphics[width=0.75\textwidth]{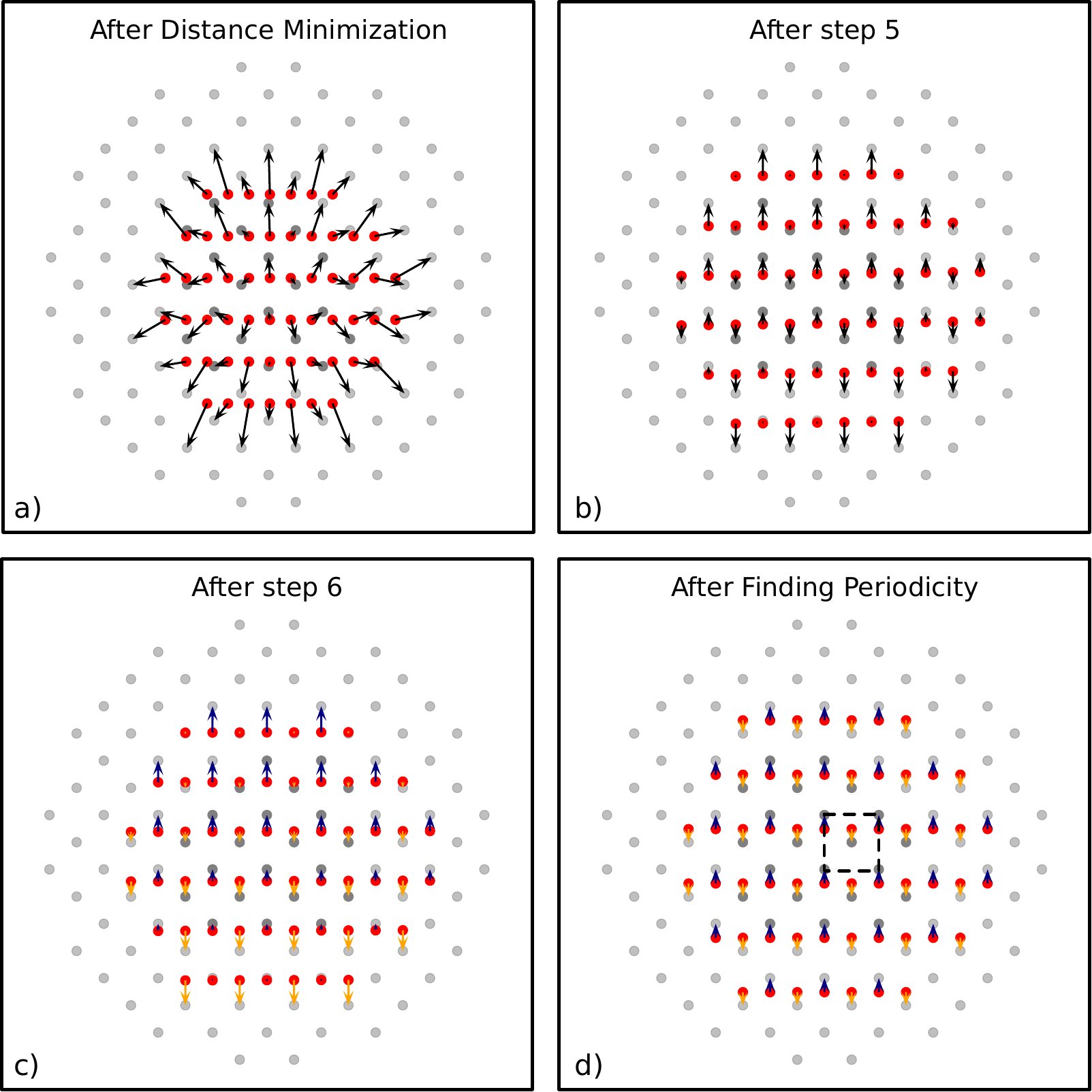}
\caption{\label{fig:post-processing}
Two-dimensional example of the post-processing steps. Each panel shows the mapping structure in red and the mapped structure in gray. The arrows represent the connections. The distance is the Euclidean distance, and the one-to-one mapping condition is enforced. (a) The system after a total distance minimization with respect to translation, rotation and mapping. (b) The system after a total distance minimization with respect to translation and linear transformation, using the mapping obtained at the previous step. (c) The system after an initial classification. Arrows of different colors represent different classes of connections. (d) The system after a class-specific standard deviation minimization with respect to translation and linear transformation, using the mapping obtained previously.}
\end{figure*}

Intuitively, one could think that the best way to apply the Hungarian algorithm is to map all $A$ atoms to all $B$ atoms. This is, in fact, problematic because there is no guarantee that the boundaries of the two spherical sections are perfectly compatible. In other words, if at the boundary of $B$ an atom needs to be mapped in the most optimal way to an atom that is outside the boundary of $A$ (it is not part of the finite section created at step 1), it will have to be mapped to some other atom of $A$ regardless. This will lead to an unwanted, exaggerated, influence of the boundaries on the final result. To prevent this from happening, the mapping structure ($B$) is made smaller than the mapped structure ($A$) by making the bottom portion of the cost matrix costless (see Fig.~\ref{fig:cost_matrix}). This means that the mapping of the outer shell of the mapping structure has no effect on the total cost and that these atoms can be considered nonexistent, they are simply placeholders. In other words, there are more agents then there are tasks; some agents will be assigned the task ``do nothing.” Thereby, since there are now less atoms in the mapping structure than in the mapped structure, each atom at the boundary of the mapping structure, can find its true counterpart in the mapped structure (provided that the mapped structure is large enough).

This inevitably leads to a new problem: there is no guarantee that all the atoms close to the center of the mapped structure will actually be mapped. In other words, some atoms of $A$ can be ``skipped" by the algorithm. When studying polymorphic transformation, this can be problematic since atoms cannot disappear when going from the initial to the final structure, i.e., every atom needs to be mapped. To avoid this problem, a very high cost can be given for mapping placeholders at the outside of the mapping structure to core atoms (important atoms) inside the mapped structure (see Fig.~\ref{fig:cost_matrix}). When an atom of the mapped structure is not mapped, it is, in fact, mapped to a placeholder in the mapping structure (it is assigned the task ``do nothing"). Therefore, imposing a very high cost to mapping core atoms to placeholders will prevent those atoms from not being mapped. Or in terms of tasks and agents: important agents cannot be assigned the task ``do nothing." Adding core atoms enforces the one-to-one mapping (bijection) between the core of the mapped structure and the corresponding subset of atoms in the mapping structure. The choice of core atoms (if any) and the relative size of the mapping structure compared to the mapped structures are also the parameters external to the algorithm (set by the user).

This concludes the distance minimization part of the algorithm (upper part in Fig.~\ref{fig:algorithm}), which leads to the optimal alignment and the atom-to-atom mapping between two structures. The result of this stage is depicted in Fig.~\ref{fig:post-processing}(a) for a simple 2D example used to illustrate various aspects of our algorithm.
%
\subsection{Finding Periodicity}
%
In the previous part of the algorithm, the distance has been minimized, and the optimal mapping has been found. The resulting $(p_{min}, Q_{min}, \vec{t}_{min})$ is only applicable to the finite portions of the two crystals that were chosen at the \textit{Tiling} step. The goal, however, is to describe the matching for the full infinite crystals, which requires finding the periodicity of the map if it exists.\\ 

We start from a vector field of \textit{connections}, that is, the vectors that go from the mapping structure ($B$) to the mapped structure ($A$) noted  $\vec{\rho}_i = \vec{a}_{i} - (Q_{min}\vec{b}_{p_{min}(i)} + \vec{t}_{min})$. The idea is to classify equivalent connections into groups, label them, and find the unit cell of the resulting ``connection crystal". To do so, the first step is to make the connections periodic. Indeed, even when the \textit{mapping} $p_{min}$ is periodic, the connections themselves are not necessarily periodic and, as already discussed, they can diverge in magnitude (see Fig.~\ref{fig:post-processing}(a)). In the example from Fig.~\ref{fig:limitations} the volumes per atom (areas in 2D) of the structures are exactly the same which implies the existence of a solution with non-diverging connections, but, in general, if the volumes are different the divergence cannot be avoided. 

The connections can be decomposed in two components: (1) a component that accounts for the difference in volumes (stretching/compressing or strain) and (2) a non-diverging, periodic component. The magnitude of the former increases as one moves away from the center of alignment. In order to reveal the periodicity, the non-diverging component needs to be isolated from the divergent one. To do so, keeping the final atom-to-atom mapping $p_{min}$ fixed, the algorithm minimizes the distance once more, but this time with respect to a linear transformation $T$ and a translation $\vec{t}$ where $det(T)$ is unrestricted (step 5, see Fig. \ref{fig:algorithm}).  An illustration of the resulting connection field after step 5 is presented in  Fig.~\ref{fig:post-processing}(b). If the structures were infinite, minimizing the distance with respect to $T$ would naturally eliminate the diverging component of the connections by making the volume per atom the same in both structures i.e., $det(T) = det(C_A)/det(C_B)$. However, since, in practice, the structures are finite, the condition on $det(T)$ is not exactly fulfilled and the connections are not yet fully periodic.     

To address this problem, the algorithm proceeds to an initial coarse classification of the connections. It is done by placing the connection vectors in different groups with respect to their norm and orientation according to a certain tolerance factor (analogous to bins when making a histogram). On Fig.~\ref{fig:post-processing}, from panel (b) to panel (c), the connections are separated into two groups: blue, pointing up and orange, pointing down. The algorithm then proceeds to making the connections in each group as similar as possible  to each other (in norms and directions) by applying an additional linear transformation to the mapping structure in order to correct the finite size effects introduced at the previous step (step 5). This is done simultaneously for all classes of connections where instead of minimizing the distance function, the algorithm minimizes the class-specific standard deviation (STD) of the connections. This step is represented mathematically by the following equation:
\begin{align}
        \vec{t}, T = \underset{\vec{t}', T'}{\text{argmin}} \sum_{i}^{N} \left(\vec{\rho}_i - \frac{\sum \limits_{j \in \Omega_i} \vec{\rho}_{j}}{\left|\Omega_i\right|}\right)^2 \label{eq:csstd}
\end{align}
where
\begin{align*}
        \vec{\rho}_i = \vec{a}_i - (T'\vec{b}_{p_{min}(i)} + \vec{t}'),
\end{align*}
and $\Omega_i$ is the class that contains $i$ and $\left|\Omega_i\right|$ denotes the number of elements in that class. The quantity to minimize in eq.~\eqref{eq:csstd} is simply a standard deviation with respect to the mean of each class. The classification (step 6) and the minimization of the STD (step 7) are repeated reducing the classification tolerance iteratively until, the STD is equal to zero. This is what we call the Classification Loop on Fig.~\ref{fig:algorithm}. In practice, making connections of each class exactly identical eliminates the remaining diverging component, which is confirmed by verifying that $det(T) = det(C_A)/det(C_B)$.  After this step, the connections are perfectly periodic, and reflect periodicity in the mapping (which has remained the same) as shown in Fig.~\ref{fig:post-processing}(d).

Using the classification of the connections, we can simply proceed as if we were to find the unit cell of a crystal made of connection vectors (instead of atoms). This structure can be described like any other crystal structure by $D=\{ C_D , P_D, L_D\}$, but in this case $L_D$ is a list of labels that indicates the atomic specie and the class of connection (e.g., blue or yellow on Fig.~\ref{fig:post-processing}(d)). The primitive cell of that structure is the scale of the matching and also an alternative unit cell $C_A'$ of structure $A$ and, consequently, also determines an alternative unit cell $C_B'$ of $B$.  $D$ and $T$ are the final results of the algorithm. They have the following properties:
\begin{subequations}
\begin{align}
    C_D &= C_A' = TC_{B}' \\
    P_{D}  &= TP_{B}' \label{eq:trans} \\
    P_D &= P_{A}' - V_D \label{eq:initial} \\
    L_D & = L_A' = L_B' ,
\end{align}
\end{subequations}
where A' and B' are alternate representations of A and B; in general they are \text{not} the ones that were input initially. $V_D$ is a matrix whose columns are the connection vectors associated with each atomic position. It can easily be constructed from $L_D$.

A full implementation of our algorithm is available online (see section \ref{sec:code}).
%
\section{Applications}
\subsection{Solid-Solid Phase Transformations}
%
To find transformation pathways using our algorithm we set the distance function to be the Euclidean distance between the atoms, we set $\varepsilon=0$ such that no amount of deformation is allowed during the distance minimization step (upper part of Fig.~\ref{fig:algorithm}) and we enforce the one-to-one mapping condition. The algorithm therefore finds the transformation  for which the total distance traveled by all the atoms to go from the initial to the final structure is minimal. The ``connection vectors" (arrows) represent the displacements of the atoms during the transition.  The output from the algorithm, $D$ and $T$, can be used to fully describe the system at any state along the transition path. 
\begin{table*}
\begin{ruledtabular}
\caption{\label{tab:validation}Result Summary of Solid-Solid Transformations. The names of the initial and final structures are given together with their chemical composition from which the lattice parameters are taken. We also provide the space group assignment for the initial, lowest symmetry intermediate and the final structures. The last column indicates whether the pathway found by our algorithm agrees with those discussed in the literature. For mechanisms that involve slipping processes, the information about the underlying mechanism (without slipping) is specified in parentheses.}
\begin{tabular}{cccccccccc}
Transformation         & Chemical comp. & \multicolumn{5}{c}{Space groups}               &     Previously reported           \\
                       &                & Initial & & Lowest Sym. Intermediate & & Final &    (without slipping)  \\
\colrule
                                 &                  &                         &                         &                         &                       &                & \\ 
HCP to BCC             & Ti             & $P6_3/mmc$    & $\rightarrow$   & Cmcm             & $\rightarrow$  & Im-3m     & Yes~\cite{burgers_1934, masuda_2004, stevanovic2018} \\
Graphite to Diamond    & C              & $P6_3/mmc$    & $\rightarrow$   & C2/m             & $\rightarrow$  & Fd-3m     & Yes~\cite{khaliullin_2011, xiao_2012, stevanovic2018} \\
FCC to BCC             & Fe             & Fm-3m         & $\rightarrow$   & $P2_1m$ (I4/mmm) & $\rightarrow$  & Im-3m     & No~(Yes~\cite{bain_1924, nishiyama_2012, stevanovic2018}) \\
Rocksalt to CsCl-type  & CsCl           & Fm-3m         & $\rightarrow$   & Pc (Pmmn)        & $\rightarrow$  & Pm-3m     & No~(Yes~\cite{capillas_2007, watanabe_1977, stevanovic2018}) \\
Roscksalt to Wurtzite  & ZnO            & Fm-3m         & $\rightarrow$   & $P3_1$           & $\rightarrow$  & $P6_3mc$  & No \\
Rocksalt to Zincblende & SiC            & Fm-3m         & $\rightarrow$   & R3m              & $\rightarrow$  & F-43m     & Yes~\cite{capillas_2007, blanco_2000}    
\end{tabular}
\end{ruledtabular}
\end{table*}  

We have tested our algorithm on several well-studied transformations. Table \ref{tab:validation} summarizes the results. For hexagonal close-packed (HCP) to body centered cubic (BCC), Graphite to Diamond and Rocksalt to Zincblende, we find pathways that have been previously reported in literature \cite{burgers_1934, masuda_2004, stevanovic2018,khaliullin_2011, xiao_2012, capillas_2007, blanco_2000}. The symmetries of the intermediate structures are exactly the same. For the other three transformations (face centered cubic (FCC) to BCC, Rocksalt to CsCl-type, Roscksalt to Wurtzite), we find new pathways that have not been reported yet. 

In the case of FCC to BCC and Rocksalt to CsCl-type, the newly found pathways involve a slipping process. It has the effect of reducing the total distance between the atoms. By preventing slipping from happening we find exactly the same transformation mechanisms (indicated in parentheses in Table \ref{tab:validation}) that have been reported before. All of the mechanisms that are discussed in literature for these transformations have been derived using periodic boundary conditions; hence, they all suffer from the problem from Fig.~\ref{fig:limitations}.  For a large number of atoms, they inevitably lead to the greater total travel distances than the same mechanisms with the added slipping process (see Fig.~\ref{fig:old_vs_new} for FCC to BCC). As we have mentioned before, the component of the displacement of the atoms (or connections) that has the most impact on the total distance is the one associated with strain. Therefore, by minimizing distance our algorithm also minimizes strain. The slipping process appears naturally because it reduces the strain associated with the transformation. In regard to Rocksalt to Wurtzite, our new pathway does not involve a simple slipping mechanism, but a more complex process which, in turn, also leads to a shorter travel distance and consequently smaller principal strains than the path with symmetry $Cmc2_1$ reported in Refs. \cite{capillas_2007, sowa_2001, stevanovic2018} (see Fig.~\ref{fig:old_vs_new}). These effects are not present in the HCP to BCC, Graphite to Diamond and Rocksalt to Zincblende, for which our algorithm agrees with the mechanisms commonly discussed in the literature, because these mechanisms already minimize the strain.

These considerations show that: (1) we have reached our goal of creating an algorithm that finds the true path of minimal distance since we either find known pathways or new pathways of shorter total distance and that (2) the result from our algorithm can not only be used as a starting point for ssNEB, but it can also be interpreted directly to explain certain features of the transformation.
\begin{figure*}
\includegraphics[width=0.9\linewidth]{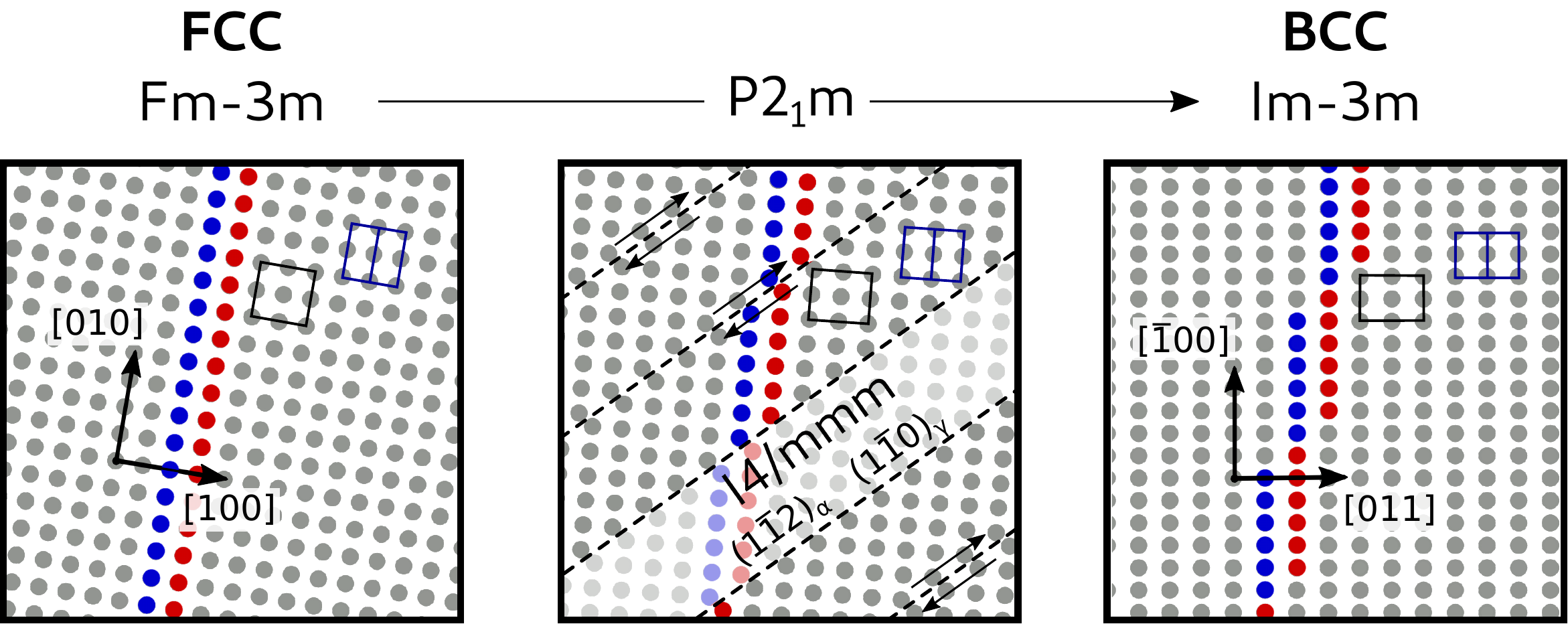}
\caption{\label{fig:martensitic}
Modeled martensitic transformation. The panel on the left represents the initial FCC structure viewed in the [001] direction. The middle panel shows the structure along the transformation. The right panel shows the structure in the final BCC structure from the $[1\overline{1}0]$ direction. In each panel, the BCC conventional cell is represented in blue and the FCC conventional cell in black. The red and blue rows of atoms are guides to help visualize the slipping process.}
\end{figure*}

For example, let us analyze in more detail the FCC to BCC transformation in iron also known as the martensitic transformation. The martensitic transformation is the diffusion-less transformation of steel from the cubic face centered (FCC) austenite ($\gamma$) phase to the body-centered cubic (BCC) or body-centered tetragonal (BCT) martensite ($\alpha$). For simplicity, we considered the transition of pure iron from FCC to BCC. For austenite, we used a lattice parameter of $a_{\gamma}=0.3585$\r{A} and we defined the lattice parameter in martensite as $a_{\alpha}=\sqrt{\frac{2}{3}}a_{\gamma}=2.2927$\r{A} such that the closed pack directions have the same atomic density for both structures. 

Figure \ref{fig:martensitic} shows the martensitic transformation found by our algorithm. The transformation consists of a main shear of the $(1\overline{1}2)_\alpha$ planes in the $[\overline{1}11]_\alpha$ direction with slip planes every six layers. Between the slip planes, the intermediate structure has the I4/mmm space group which correspond to a Bain distortion accompanied by a rotation. Transformation mechanisms that involve a rotated Bain deformation have been widely theorized \cite{bain_1924, nishiyama_2012, zhang_kelly_2005}. As we mentioned, the occurrence of slip planes can be explained by the fact that they greatly reduce the strain necessary to carry out the transformation. The principal strains for our new mechanism are -5.7\%, 0\% and 15.5\%, whereas they would be -18.4\%, 15.5\% and 15.5\% without the occurrence of slip planes (Bain distortion). This slipping process is often used in the context of the Phenomenological Theory of Martensitic Transformation \cite{bowles1951crystallographic, bowles1954crystallography, mackenzie1954crystallography, Wechsler1953, Wechsler1960} to explain the occurrence of striations along the $(1\overline{1}2)_\alpha$; our algorithm finds it naturally by minimizing distance. 

The number of layers between the slip planes depends on the ratio between the parameters of the initial and final structure. The connection structure is composed of 6 atoms which means that the transformation occurs at a scale that corresponds to 6 primitive cells of the two end structures (they have the same number of atoms). Once again, our algorithm behaves as expected by finding a transformation that reduces the total travel distance--and thus the strain--and by being able to find transitions that occur on a larger scale. A more detailed analysis of our results for the martensitic transformation will be published elsewhere \cite{therrien_2020}.

In this study of the martensitic transformation, we used 1000 random initialization steps (Random Initialization Loop), a mapping structure of 180 atoms and a mapped structure of 600 atoms with the one-to-one mapping enforced. Those parameters ensure that the finite portion of the crystal is much larger than the connection cell and that the global minimum is reached. The calculation took 6~min~41~s on a 36-core Intel Xeon Gold 6154 (3.00~GHz) node and 2~h~12~min~44~s on a 4-core Intel Core i7-8550U (1.80~GHz) Laptop. Using the same parameters, the minimization for HCP to BCC, Rocksalt to CsCl-type and Rocksalt to Zincblende were done in 7~min~31~s, 2~min~47~s and 1~min~14~s respectively on the computing node. For the transition from graphite to diamond which involves a large change in specific volumes, we used a mapping structure of 160 atoms and a much larger mapped structure of 1600 atoms. This was done to ensure that there was a sufficient number of graphite layers in the mapped structure. We also increased the number of random initial steps to 3000 in order to find the global minimum. This calculation was completed in 29~h~54~min on the compute node. Similarly, for the transition from Rocksalt to Wurtzite we used a mapping structure of 600 atoms and a mapped structure of 2000 atoms with 2000 initial steps; the calculation was completed in 5~h~14~min. The limiting factor in the distance minimization is the resolution of the assignment problem, therefore the complexity of the mapping dictates the total execution time. Future development may involve the use of an implementation of the Hungarian algorithm parallelized for graphics processing units (GPUs) \cite{date2016gpu}.

%
\subsection{Semi-Coherent Interfaces}
%
Next, we illustrate how our algorithm can also be used to find the structures of semi-coherent interfaces between different materials. In the examples that follow, we consider only the terminating planes in each structure. Therefore, in our algorithm the two structures are modeled as large disk-like 2D sections of the terminating planes (instead of spheres in 3D). For demonstration purposes, here, the plane directions and terminating layers are taken from experiment. Each connection between an atom from the mapped structure and an atom from the mapping structure represents a chemical bond. Atoms no longer have to be mapped to atoms of the same specie, they are mapped according to chemistry rules that determine which types of atoms from one structure will bond to which type of atoms from the other structure e.g., Zr atoms bond with Ni atoms. These rules need to be known in advance. Since connections now represent chemical bonds, for the distance metric in equation \ref{eq:min}, we use the Lennard-Jones potential:
\begin{align}
\text{d}(\vec{a},\vec{b}) = \sigma\left(\frac{r^{12}}{||\vec{a}-\vec{b}||^{12}} - 2\frac{r^{6}}{||\vec{a}-\vec{b}||^{6}}\right) \label{eq:LJ}
\end{align}
where $\sigma$ denotes the potential strength and $r$ the equilibrium radius. We use this potential because it is a mathematically simple representation of the general shape of the potential between 2 atoms. In our model, atoms are bonded with at most 1 atom of the other phase. In other words, we assume that the bond with the closest neighboring atom of the other phase is the strongest and most consequential in terms of energy and alignment. Since we are only interested in the optimal alignment--we are not trying to predict the interfacial energy, the strength of the potential $\sigma$ is not important and it is set to 1. Thus, the potential has only one parameter: the equilibrium radius $r$. It can be set based on physical or experimental arguments.
Moreover, there is no need to enforce that each and every atom of both structures form a bond. In the case of semi-coherent interfaces for example, the lattice constant of the two materials can be very different such that only a fraction of the atoms at the interfaces will form bonds. Because, in our model, an atom can form at most 1 bond (1 or 0 bond), there cannot be more bonds per unit area than there are atoms per unit area in the structure that is the least dense. Since, in general we wish to maximize the number of bonds per unit area, all the atoms of the least dense phase need to form a bond. This is done by setting the denser structure as the mapped structure and by setting the fraction of core atoms in the cost matrix to 0 such that the one-to-one mapping is not enforced. In fact, in this case, we take advantage of the fact that certain atom will naturally be ``skipped" when making the mapping structure smaller than the mapped structure.
Finally, during the distance minimization step (in this case the distance is defined by equation \ref{eq:LJ}), the structures may be slightly strained in-plane near the interface in order to maximize the bonding energy. Therefore, we usually set $\varepsilon$ to a value between 3-8\%.
\begin{figure*}
\includegraphics[width=0.85\linewidth]{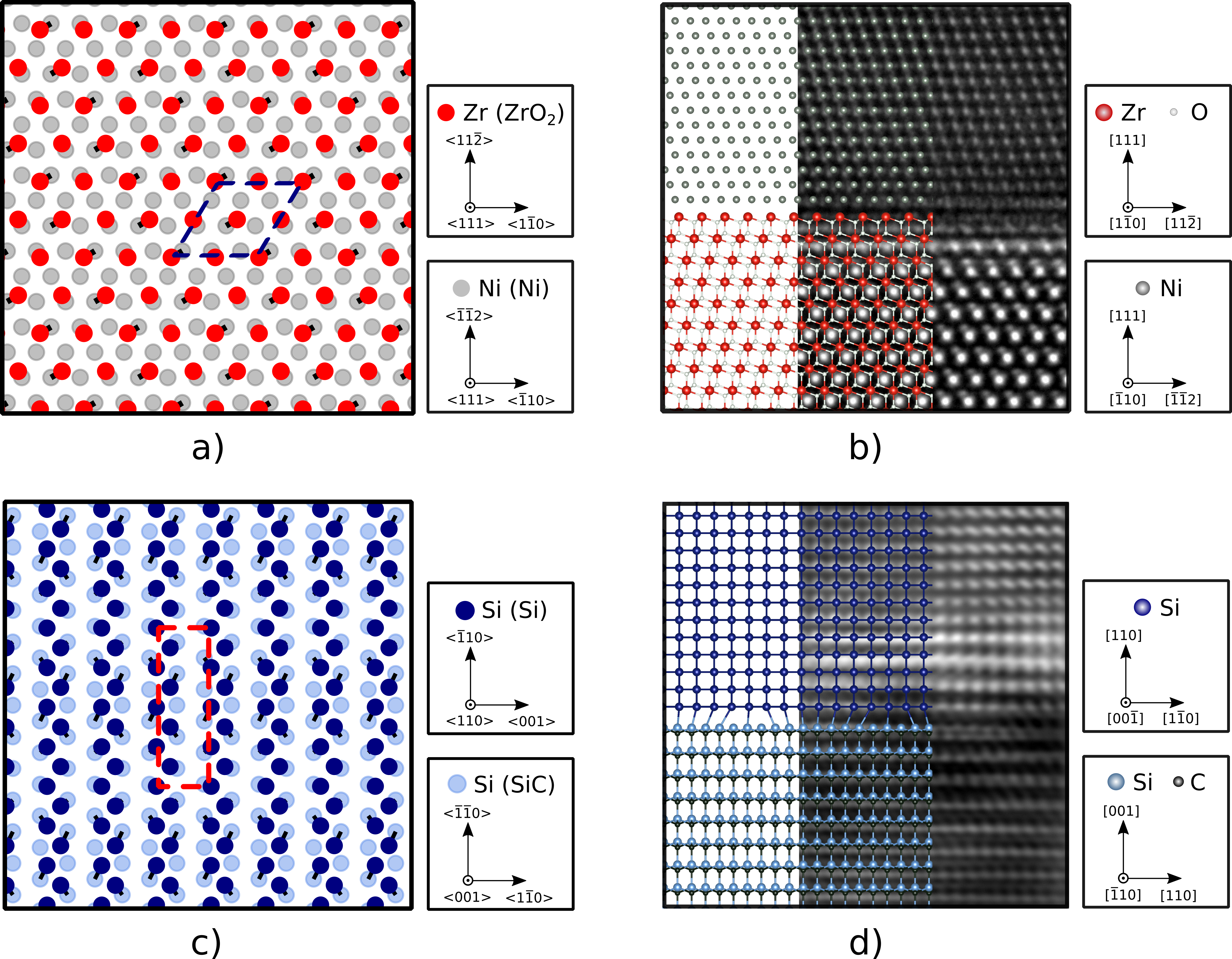}
\caption{\label{fig:interfaces}
Examples of interface models from our algorithm and comparison with experimental results. (a) In-plane (111) view of the Ni on $\text{ZrO}_2$ interface model. The repeating matching pattern is highlighted by a blue dashed parallelogram. (b) Ni$[\overline{1}10]$ projection of the model (left) partially overlaid on an HRTEM micrograph of the same projection taken from Ref.~\cite{nahor_2016}. The atomic columns are placed on the micrograph according to the simulations in Ref \cite{nahor_2016}. (c) In-plane Si(110) view of the Si on SiC interface model. The repeating matching pattern is highlighted by a red dashed parallelogram. (d) Si$[00\overline{1}]$ projection of the model partially overlaid on an HRTEM micrograph of the same projection taken from Ref.~\cite{Li_2016}. We processed the original image using the Fourier masks filtering technique and contrast enhancement. The atomic columns are placed on the micrograph such that the periodicity of our model can be easily compared to that of the micrograph.}
\end{figure*}

We used our algorithm to find the orientation between two experimentally realized interfaces.
The first system is a solid-solid interface between Ni and yttrium-stabilized zirconia (YSZ). This interface was experimentally realized and studied by Nahor et al. \cite{nahor_2016}. We used the parameters from their experiment to run our simulation. For face-centered cubic Ni, we used a lattice parameter of 3.52\r{A} and for cubic $\text{ZrO}_2$, we used a lattice parameter of 5.125\r{A} (apart from its effect on the lattice parameter, the presence of yttrium was not considered in our simulation). We specified the interfacial plane (111) for both structures and used the termination (Zr for $\text{ZrO}_2$) specified  in Ref. \cite{nahor_2016}. We set the equilibrium point of the Lennard-Jones potential $r$ to be 2.5\r{A}, because it is close to the Ni-Ni interatomic distance of 2.49\r{A}. The value of the equilibrium point is an estimate of the length of the bonds and therefore it determines the distance between the layers of the two phases. We find that its value does not have a strong incidence on the final result. We set $\varepsilon$ to 0.08 such that there can be some strain in the layers close to the surface. 

We find Ni${<}\overline{1}10{>}(111)$//$\text{ZrO}_2$ ${<}1\overline{1}0{>}(111)$ \footnote{Our algorithm does not differentiate between $\text{OR}_1$ and $\text{OR}_2$ because they are identical in-plane.} to be the optimal alignment between the two structures in accordance with the experimental observation.  Fig.~\ref{fig:interfaces}(a) shows the interface viewed in the [111] projection (from above). The algorithm finds a repeating pattern of only a few unit cells in sharp contrast with the $33\times33$ unit cells O-lattice found using the measured orientation relationship \cite{nahor_2016}. This smaller cell has the advantage of directly providing the matching modes of the interfaces: 2:3 Zr-Ni which is in accordance with the ``one dislocation every three Ni planes" observed by Nahor et al. This smaller cell is possible because the optimal result was found by allowing some strain in the interface layers. In fact, in the result shown in Fig.~\ref{fig:interfaces}, the ratio between the area of the Zr unit cells in-plane and the Ni unit cells in-plane is increased by 6.12\%. In other words, there is tension in the YSZ side and compression on the Ni side also in agreement with the observation of Nahor et al. Not only does our algorithm correctly reproduce the measured orientation relationship solely using the in-plane lattice parameters, but it also provides the matching mode and the general direction of the strain (tension or compression) in the layers near the interface. A projection of our model along the Ni$[\overline{1}10]$ direction with the aforementioned in-plane strain is presented in Fig.~\ref{fig:interfaces}(b) in comparison with an HRTEM micrograph of the same projection. The model is in very good agreement with the experimental result.

The second system is the solid-solid interface between Si and SiC. This interface was experimentally realized and studied by Li et al. \cite{Li_2016}, once again, we used the parameters from their experiment to run our simulation. For diamond Si we used a lattice parameter of 5.43\r{A} and for hexagonal close-packed (HCP) 6H-SiC we used a lattice parameter of 3.08\r{A}. We set $\varepsilon$ to 0.08. As specified in Ref. \cite{Li_2016}, we matched the (110) plane of Si with the Si terminated (001) plane of SiC. We set the equilibrium point of the Lennard-Jones potential ($r$) to be 2.5\r(A), because it is close to the Si-Si interatomic distance of 2.35\r(A).

We find the Si${<}1\overline{1}0{>}(110)$//6H-SiC${<}110{>}(001)$ orientation to be the optimal alignment between the two structures in accordance with the experimental observation. Fig.~\ref{fig:interfaces}(c) shows the interface viewed in the Si[110] projection (from above). Once again, our periodic pattern is in accordance with the observed 4:5 Si to SiC matching mode in the Si$[1\overline{1}1]$/SiC$[110]$ direction. In addition, the algorithm finds a 1.68\% increase in the ratio between the in-plane Si(SiC) cell and the in-plane Si(Si) cell. In other words, the 6H-SiC structure is stretched and/or the Si structure is compressed at the interface to obtain that ratio. This could explain the 1.84\% mismatch in the Si[001] direction and the 0.26\% residual mismatch in the Si$[\overline{1}10]$ direction noted by Li et al. A projection of our model along the Si$[00\overline{1}]$ direction with the aforementioned in-plane strain is presented on Fig.~\ref{fig:interfaces}(d) in comparison with an HRTEM micrograph of the same projection; the model is in very good agreement with the experimental result.

We obtained the result presented above using the same type of 36-core Intel Xeon computing nodes. For Ni//YSZ the minimum was obtained in 1~min~35~s using a mapping structure of 325 atoms and a mapped structure of 1300 atoms with 1000 initializations. For Si//SiC we used a mapping structure of 150 atoms and a mapped structure of 600 atoms with the same number of random initializations. The calculation was completed in 13~min~39~s. 
%
\section{Measuring Distance Between Crystal Structures}
%
As discussed previously, defining a rigorous Euclidean distance between crystal structures or more broadly between infinitely periodic arrays of points is a challenging task. First, the infinite dimensionality of the configuration space poses problem. This difficulty can, in principle, be avoided by scaling the metric with some function of the number of atoms $N$. As we will show in this section, it is actually not possible as the dependence on the number of atoms involves different powers of $N$. Secondly, and as importantly, even a finite portion of a crystal structure is not represented by a unique point in the $N$-dimensional configuration space of atomic coordinates, but by several points that reflect: (a) the permutations of atomic indices that describe the same crystal structure, and (b) the variability in the choice of the $N$-atom section of an infinitely periodic crystal. The  definition of a distance metric between two crystal structures for any fixed $N$ implies finding the two closest representative points of the two structures in the $N$-dimensional configuration space, a task that is tackled by our algorithm. In fact, once the optimal parameters $(p_{min}, Q_{min},\vec{t}_{min})$ defined in equation~\eqref{eq:min} are found, for a fixed $N$, the minimized distance may serve as a mathematical metric between periodic structures.

Let us consider the distance between two structures in the situation where the correspondence between them has already been established. Since the mapping is periodic, the two structures ($A$ and $B$) in their optimal matching can be described with cells $C_A$ and $C_B$ which both contain $m$ atoms and are optimally aligned, and with atomic positions $\{\vec{a}_i \,|\, i=1,2,\dots,m\}$ and $\{\vec{b}_i \,|\, i=1,2,\dots,m\}$ inside the cells indexed according to the optimal mapping. The shortest travel distance between the two structures with this match is: 
\begin{equation}\label{eq:dist1}
d_1 = \sum_{l=1}^{m}\sum_{i,j,k = -\frac{n}{2}}^{\frac{n}{2}}||\vec{a}_{ijkl} - \vec{b}_{ijkl}|| ,
\end{equation}
where $\vec{a}_{ijkl}$ stands for the position of an atom that belongs to the structure $A$. $\vec{a}_{ijkl}$ is a periodic image of an atom with an index $l$ located in the unit cell indexed with $ijk$. More precisely: 
\begin{equation}
    \vec{a}_{ijkl} = C_{A}\mat{i\\j\\k} + \vec{a}_l,
\end{equation}
and analogously for the structure $B$. The total number of atoms in each structure is $N=m(n+1)^3$. This distance ($d_1$) is, by construction, the $l_{1,2}$-norm \cite{ding2006r,nie2010efficient} of the $3 \times N$ matrix formed by the connection vectors. We used the same norm when posing the matching problem for phase transitions.
So far, we used the $l_{1,2}$-norm because it represents the sum of the distances traveled by all the atoms during the transition, but one could also be interested in computing the Frobenius norm of that matrix i.e., the $l_2$-norm of the vector joining the two structures in configuration space. It is given by:
\begin{align}
d_2 = \sqrt{\sum_{l=1}^{m}\sum_{i,j,k = -\frac{n}{2}}^{\frac{n}{2}}||\vec{a}_{ijkl} - \vec{b}_{ijkl}||^2} . \label{eq:dist2}
\end{align}
Adding a $1/N$ factor inside the square root in front of the summation gives the root mean square distance (RMSD). It is important to note that, for finite $N$, the set of optimal parameters $(p_{min}, Q_{min},\vec{t}_{min})$ is not necessarily the same for $d_1$ and $d_2$ 
When they are optimized, both $d_1$ and $d_2$ fulfill the 4 requirements of a metric: 
 \begin{enumerate}
 \item $d(A,B)\geq0$, 
 \item $d(A,B) = 0 \iff A=B$, 
 \item $d(A,B) = d(B,A)$ and 
 \item $d(A,C) \leq d(A,B) + d(B,C)$. 
 \end{enumerate}
 The first 3 criteria follow trivially from the properties of the $l_2$-norm applied to the connection vectors. The fourth criterion follows from the fact that both $d_{1}(A,C)$ and $d_{2}(A,C)$ are defined respective to their $(p_{min}, Q_{min},\vec{t}_{min})$ and represent the shortest distance between A and C.
The problem is that both $d_1$ and $d_2$ depend on $N$ and in order to compare actual structures, we either need: (1) to derive quantities from $d_1$ and $d_2$ that are independent of $N$, or (2) find a way to compare distances in the limit where $N\to\infty$.
In both cases, the first step is to derive the dependence of $d_1$ and $d_2$ on $N$.
%
\subsection{Size dependence}  
%
The case of $d_2$ can be derived analytically so we will use it for demonstration purposes.
Let us first define $\vec{\rho}_l = \vec{a}_l - \vec{b}_l$ and $C' = (C_A - C_B)$ whose columns are $\{\vec{c}_\nu \, | \, \nu=1,2,3 \}$. From equation~\eqref{eq:dist2}, we can now write:
\begin{align}
d_2^2 = \sum_{l=1}^{m}\sum_{i,j,k = -\frac{n}{2}}^{\frac{n}{2}}||C'\mat{i\\j\\k} + \vec{\rho}_l ||^2 .
\end{align}
If the cells are exactly identical, then $C'=0$ and it follows that:
\begin{align}
d_2^2 = (n+1)^3\sum_{l=1}^{m} ||\vec{\rho}_l||^2 = K_2N , \label{eq:dist2_lin}
\end{align}
where
\begin{equation} \label{eq:K22}
 K_2 = \frac{1}{m}\sum_{l=1}^{m} ||\vec{\rho}_l||^2.
\end{equation}
If the cells are not identical, $C'$ is an invertible matrix and we can write:
\begin{align}
d_2^2 =& \sum_{l=1}^{m}\sum_{i,j,k = -\frac{n}{2}}^{\frac{n}{2}}||C'\mat{i+{\rho'}_l^{(1)}\\j+{\rho'}_l^{(2)}\\k+{\rho'}_l^{(3)}}||^2 ,
\end{align}
where $\{{\rho'}_l^{(\nu)}\}$ are the elements of $C^{-1}\vec{\rho}_l$. Then, using the fact that the vector norm squared is equivalent to the inner product of the vector with itself:
\begin{align}
  \begin{split}
    d_2^2 &=  \sum_{l=1}^{m}\sum_{i,j,k = -\frac{n}{2}}^{\frac{n}{2}} (I\vec{c}_1 + J\vec{c}_2 + K\vec{c}_3)\cdot(I\vec{c}_1 + J\vec{c}_2 + K\vec{c}_3) \\
&= T_{11} + T_{22} + T_{33} + 2T_{12} + 2T_{13} + 2T_{23} ,
  \end{split} \label{eq:dist2_terms}
\end{align}
where
\begin{equation*}
I = i+{\rho'}_l^{(1)},\, J = j+{\rho'}_l^{(2)},\, K = k+{\rho'}_l^{(3)} . 
\end{equation*}
This gives 6 terms, that can be broken down into two cases. First:
\begin{align}
\begin{split}
T_{11} &= \sum_{l=1}^{m}\sum_{i,j,k = -\frac{n}{2}}^{\frac{n}{2}} (i+{\rho'}_l^{(1)})^2||\vec{c}_1||^2\\
&= \frac{(n+1)^3n(n+2)m}{12}||c_1||^2 \\
& \phantom{aaaaa} + (n+1)^3\sum_{l=1}^{m} ({\rho'}_l^{(1)})^2||c_1||^2
\end{split}
\end{align}
(the $T_{22}$ and $T_{33}$ cases are similar), and second:
\begin{align}
\begin{split}
T_{12} =& \sum_{l=1}^{m}\sum_{i,j,k = -\frac{n}{2}}^{\frac{n}{2}} (i+{\rho'}_l^{(1)})(j+{\rho'}_l^{(2)})\vec{c}_1\cdot\vec{c}_2\\
=& (n+1)^3\sum_{l=1}^{m} {\rho'}_l^{(1)}{\rho'}_l^{(2)}\vec{c}_1\cdot\vec{c}_2
\end{split}
\end{align}
(the $T_{13}$ and $T_{23}$ cases are similar).
Finally, equation~\eqref{eq:dist2_terms} becomes:
\begin{align}
\begin{split}
d_2^2 = \frac{(n+1)^3n(n+2)m}{12}(||\vec{c}_1||^2+||\vec{c}_2||^2+||\vec{c}_3||^2) \\+  (n+1)^3\sum_{l=1}^{m}\sum_{\nu}^3\sum_{\mu}^3 {\rho'}_l^{(\nu)}{\rho'}_l^{(\mu)}\vec{c}_\nu\cdot\vec{c}_\mu .
\end{split}
\end{align}
Substituting $n=\left(\frac{N}{m}\right)^{\frac{1}{3}} - 1$, and regrouping the constants, we get the following relation:
\begin{align}
d_2^2 = G_2N^{\frac{5}{3}} + K_2N \label{eq:dist2_fin} ,
\end{align}
where
\begin{align}
    G_2 =& \frac{(||\vec{c}_1||^2+||\vec{c}_2||^2+||\vec{c}_3||^2)}{12m^\frac{2}{3}} \, ,  \label{eq:K1} \\
    \begin{split}  \label{eq:K2}
    K_2 =& \frac{1}{m}\sum_{l=1}^{m}\sum_{\nu}^3\sum_{\mu}^3 {\rho'}_l^{(\nu)}{\rho'}_l^{(\mu)}\vec{c}_\nu\cdot\vec{c}_\mu \\
    & \quad - \frac{1}{12}(||\vec{c}_1||^2+||\vec{c}_2||^2+||\vec{c}_3||^2)
    \end{split}
\end{align}
Note that neither $G_2$ nor $K_2$ are dependant on the number of atoms $N$.
It is immediately evident that dividing by $N$ to any power will not result in a size independent (scaled) metric. Therefore, the RMSD (i.e., $\frac{d_2^2}{N}$) depends on the size of the system. It cannot be applied to atoms inside a unit cell (or any finite portion of the crystal) to measure distances between \textit{periodic} structures. $G_2$ is solely dependent on the difference between $C_A$ and $C_B$; if the cells remain invariant during the transformation, $G_2=0$ and equation~\eqref{eq:dist2_fin} simplifies to equation~\eqref{eq:dist2_lin}. We can say that $G_2$ is associated with the change in unit cells whereas $K_2$, the linear term, is associated with the displacements inside the cell.

Finding the relation for the $l_{1,2}$-norm is slightly more involved (see Appendix~\ref{sec:proof}) but we find a similar relationship:
\begin{align}
d_1 = G_1N^{\frac{4}{3}} + K_1N + \mathcal{O}(N) , \label{eq:dist1_fin}
\end{align}
where $G_1\geq0$. Once again there is no trivial way to make the distance an intensive quantity because of the presence of a non-linear term associated with the distortion in the unit cell.

Since we are only interested in $d_1$ and $d_2$ in the limit of $N\to\infty$, one might think that the leading terms in $N$, $G_2$ and $G_1$ respectively, could directly serve as metrics. However, $G$ ($G_1$ or  $G_2$) does not fulfill the second criteria of a metric since there could, in principle, exist a transformation that consists of a pure reorganization of the atoms where $G=0$ even though the end structures are different. The only way to define a proper metric is to use \textit{all} the parameters in equations~\eqref{eq:dist2_fin} or \eqref{eq:dist1_fin} depending on the particular choice. Providing that those parameters are known, the most straightforward approach, is to compare distances in the limit. 
However, since comparing functions in the limit can be tedious and not convenient for computation, we also defined a metric function that uses both parameters ($G$ and $K$), it is presented in Appendix~\ref{sec:metric}. 
%
\subsection{Practical Use}
%
We chose to study solid-solid phases transitions using $d_1$ because it represents the sum of the Euclidean distances travelled by all the atoms in the structure. In that context $d_1$ can also be seen as the true Euclidean distance between two structures and it can be used to measure distances between them. We were not able to find a general closed form for $G_1$, therefore it is not possible to obtain it directly from the optimal sets $\{C_A, P_A, L_A\}$ and $\{C_B, P_B, L_B\}$ that are found by our algorithm. However, we were able to find the general dependence of $d_1$ on $N$.  Using the optimal mapping to compute the distance $d_1$ at different sizes, we can show that the distance indeed grows according to equation~\eqref{eq:dist1_fin}.
\begin{figure}
\includegraphics[width=1.0\linewidth]{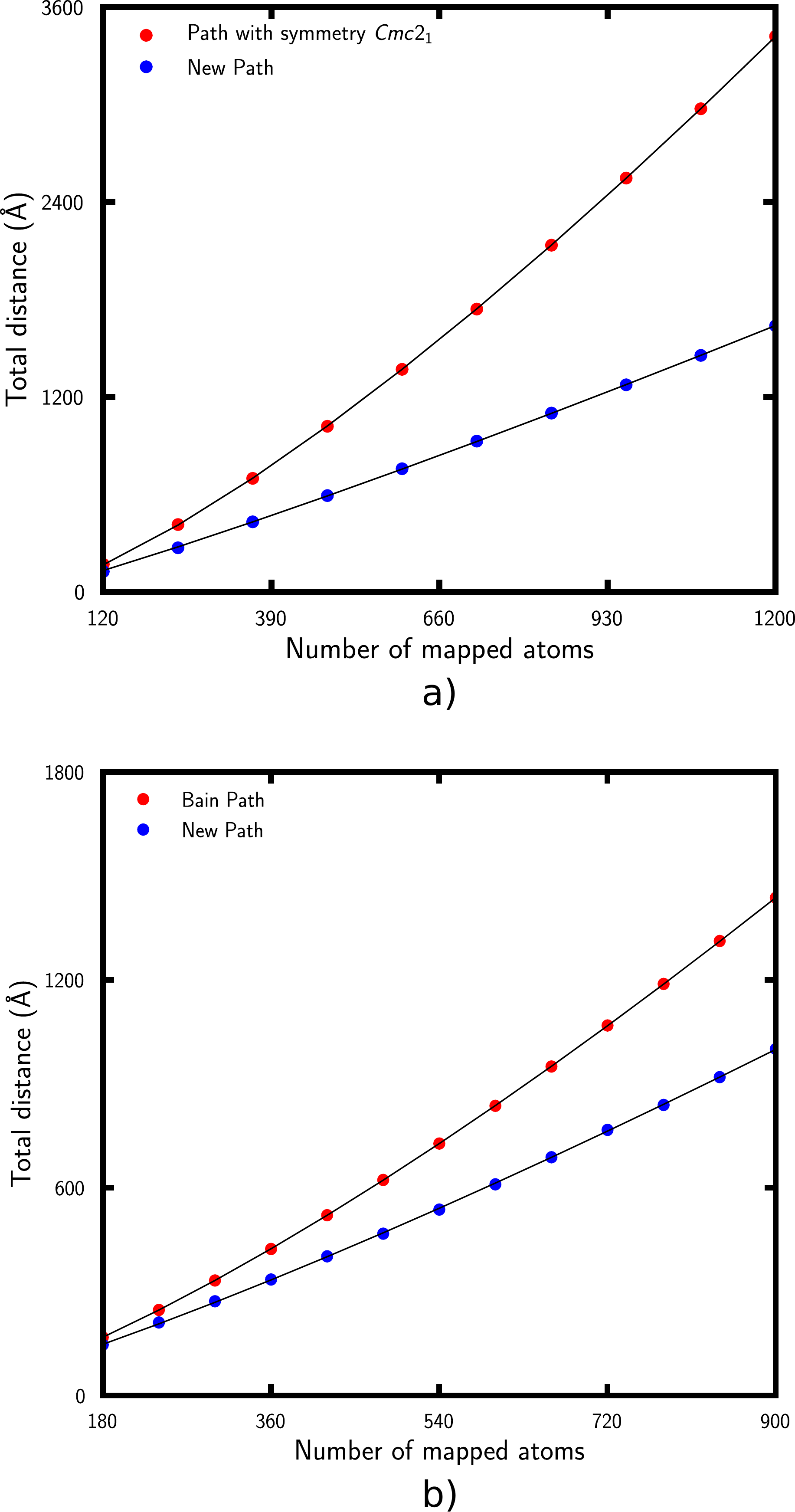}
\caption{\label{fig:old_vs_new}
Total distance traveled by all the atoms in the system as a function of the number of mapped atoms for a) the transition of ZnO from Wurtzite to Rocksalt, and b) the transition of iron from Body Centered Cubic to Face Centered Cubic. Simulated data is shown as dots, fits are represented as black lines.}
\end{figure}

Fig.~\ref{fig:old_vs_new} shows in blue the dependence of the total travelled distance $d_1$ on the number of atoms $N$, for 2 of the 6 transitions presented in Table \ref{tab:validation}. This distance is compared with the one that corresponds to the pathways previously discussed in the literature that do not involve slipping and are hence suboptimal with respect to minimizing the $d_1$. Panel (a) compares $d_1$ norms for these two pathways for the transition of ZnO from Rocksalt to Wurtzite. The red points correspond to the distance for the path with symmetry $Cmc2_1$ reported by Refs. \cite{capillas_2007, sowa_2001, stevanovic2018}. As already noted, our new pathway (in blue) produces a shorter travel distance that grows slower with  $N$. 

Fitting equation~\eqref{eq:dist1_fin} confirms the derived dependence of $d_1$ on $N$ and shows that $G_1=0.05\text{\AA}$ for the new pathway is much smaller than $G_1=0.26\text{\AA}$ for the one found in literature. The fitting curves are shown as black lines; they overlap the simulated data almost perfectly. Similarly, the red points in panel (b) correspond to the Bain deformation of Iron from FCC to BCC. Again, the new pathway has a shorter travel distance. In this case, $G_1=0.07\text{\AA}$ for the new path and $0.17\text{\AA}$ for the Bain path. Equation~\eqref{eq:dist1_fin} fits both data sets very well.
Thus, in practice, one can use our algorithm to fit $G_1$ and $K_1$ and use them to compare distances in the limit or using our metric presented in Appendix~\ref{sec:metric}. 

On the other hand, if one wishes to use $d_2$ to compare periodic crystal structures, our algorithm, can be used to find the optimal sets $\{C_A, P_A, L_A\}$ and $\{C_B, P_B, L_B\}$  by setting the distance function as the square of the euclidean norm. Then, one can use the closed form for $G_2$ and $K_2$ provided in equation~\eqref{eq:K1}, equation~\eqref{eq:K2} and equation~\eqref{eq:K22} to compare distances in the limit or using our metric presented in Appendix~\ref{sec:metric}.

In sum, in this section, we showed that distances such as the RMSD cannot be used directly to compare periodic structure because they depend on the size of the system. Instead, we established the dependence on $N$ for two metrics $d_1$ and $d_2$ and we showed how they can be used, in the limit, to quantify the similarity between structures.
%
\section{Conclusions}
%
In this work, we formulated the matching of two different crystal structures as an optimization problem and described our algorithmic solution to it. The methodology that we developed, inspired by the Iterative Closest Point, is constructed to work on large and finite portions of the two crystal structures rather than on some choice of a periodic unit. It consists of a sequential minimization of a given distance function with respect to the permutations of atomic indices and linear transformations (rotations and translations) of the atomic positions. The sequence is repeated iteratively until the convergence is achieved. After the optimal alignment of the structures and the optimal atom-to-atom map are found, our algorithm analyzes the result and retrieves the periodicity in the match. This last step ensures that the boundaries have no influence on the final result. 

We presented two different implementations of our algorithm tailored for their respective class of applications. First, we demonstrated our algorithm's relevance when studying phase transformations by examining six well-studied transformations. In each case, we either confirmed an existing mechanism or uncovered a new lower-strained pathway. In particular, for the martensitic transformation,  we found a new modified version of the Bain path that does not require large expansion along certain crystallographic directions. Then, we showed that, starting solely from the in-plane lattice parameters, our algorithm was capable of reproducing the features of experimental interface structures such as their orientation relationships, matching modes and strain directions for two case examples: Ni on YSZ and Si on SiC.

Finally, we analyzed and discussed a practical formulation of a rigorous distance metric between crystal structures that can be used to assess their Euclidean ``closeness''.

\section{Data Availability} \label{sec:code}
A full implementation of our algorithm is available via \textit{Github} at https://github.com/ftherrien/p2ptrans.
All the parameters necessary to reproduce the examples presented in this study are  included in this article. All other relevant data is available from the corresponding author on request.

\section{Acknowledgements}
This work was supported by the Center for the Next Generation of Materials by Design, an Energy Frontier Research Center funded by the U.S. Department of Energy, Office of Science, Basic Energy Sciences; and by the National Science Foundation grant No. DMR-1945010. The research was performed using computational resources sponsored by the Department of Energy's Office of Energy Efficiency and Renewable Energy, located at the National Renewable Energy Laboratory.



\appendix

\section{An additional metric} \label{sec:metric}
Since using the limit might be inconvenient computationally, let's define a metric $\mathcal{D}$ as such:
\begin{align}
\mathcal{D}(A,B) = 
\begin{cases} 
      e-e^{-K} & \text{if}\; G=0 \\
      e^{G} & \text{otherwise}
\end{cases}
\end{align}
This fulfills the four criteria:
\begin{enumerate}
    \item $G\geq0$ and $K\geq0$ if $G=0$, therefore $\mathcal{D}(A,B)\geq0$
    \item $\mathcal{D}(A,B)=0 \iff G=K=0 \iff A=B$
    \item $d_{1,2}(A,B) = d_{1,2}(B,A) \implies \mathcal{D}(A,B) = \mathcal{D}(B,A)$
    \item \begin{itemize}
        \item{if $G(A,C),G(A,B),G(B,C) > 0$:}
        
        Follows from the fact that $d_1$ is a metric and $e^{G}$ is monotone
        \item{if $G(A,C),G(A,B) > 0, G(B,C) = 0$ (or $G(A,C),G(B,C) > 0, G(A,B) = 0$):}
        
        Then $C_B = C_C \implies G(A,C)=G(A,B)=G \implies e^{G} \leq e^{G} + e-e^{-K(B,C)}$   
        \item{if $G(A,B),G(B,C) > 0, G(A,C) = 0$:}
        
        $e-e^{-K(A,C)} \leq e^{G(A,B)} + e^{G(B,C)}$ since $e-e^{-K(A,C)} \leq 1$ and $e^{G(A,B)} \geq 1$
        \item{if $G(A,B),G(B,C),G(A,C) = 0$:}
        
        Follows from the fact that $d_1$ is a metric and $e-e^{-K}$ is monotone
        \item{if two of the three $G$s are equal to 0}
        
        Then $C_A = C_B = C_C$ which is equivalent to the previous case
    \end{itemize}
\end{enumerate}
\section{N-dependence of \texorpdfstring{$d_1$}{d1}} \label{sec:proof}
Once again let's define $C' = (C_A - C_B)$ whose columns are $\{\vec{c_\nu}\, | \,\nu=1,2,3\}$ and $\vec{\rho}_l = \vec{a}_l - \vec{b}_l$, we can rewrite equation~\eqref{eq:dist1}:
\begin{align}
d_1 = \sum_{l=1}^{m}\sum_{i,j,k = -\frac{n}{2}}^{\frac{n}{2}}||C'\mat{i\\j\\k} + \vec{\rho}_l|| . \label{eq:dist1_start}
\end{align}
If the cells are exactly identical, $C'=0$ and it follows that:
\begin{align}
d_1 = (c+1)^3\sum_{l=1}^{m} ||\vec{\rho}_l|| = G_1N , \label{eq:dist1_lin}
\end{align}
where
\begin{equation}
G_1 = \frac{1}{m}\sum_{l=1}^{m} ||\vec{\rho}_l||.
\end{equation}
Let's consider the simpler case where the vectors of $C_1$ and $C_2$ are orthogonal and where $\vec{\rho}_l=0$. One can show that aligning the 3 vectors is the optimal alignment $C_1$ and $C_2$ and therefore, the vectors of $C'$ are also orthogonal. Equation~\eqref{eq:dist1_start} becomes:     
\begin{align}
d_1 = \sum_{i,j,k = -\frac{n}{2}}^{\frac{n}{2}}\sqrt{i^2||\vec{c}_1||^2 + j^2||\vec{c}_2||^2 + k^2||\vec{c}_3||^2} . \label{eq:dist1_simple}
\end{align}
Let's approximate the summation using the right Riemann sum:
\begin{multline}
\sum_{i=1}^{n}f\left(a+i\frac{(b-a)}{n}\right)\frac{(b-a)}{n} \\= \int \limits_a^b f(x)\,dx + \mathcal{O}\left(\frac{M(b-a)^2}{n}\right) ,
\end{multline}
where $M=\text{max} f'(x)$. In our case, $a=-n/2$, $b=n/2$ and:
\begin{align}
f(x) &= \sqrt{x^2||\vec{c}_1||^2 + j^2||\vec{c}_2||^2 + k^2||\vec{c}_3||^2} \\
f'(x) &= \frac{2 x||\vec{c}_1||^2}{2 \sqrt{x^2||\vec{c}_1||^2 + j^2||\vec{c}_2||^2 + k^2||\vec{c}_3||^2}} \leq ||\vec{c}_1|| .
\end{align}
Replacing in equation~\eqref{eq:dist1_simple}, we get:
\begin{align}
\begin{split}
d_1 = \sum_{j,k = -\frac{n}{2}}^{\frac{n}{2}} \int \limits_{-n/2}^{n/2} \sqrt{x^2||\vec{c}_1||^2 + j^2||\vec{c}_2||^2 + k^2||\vec{c}_3||^2}\,dx \\+ \mathcal{O}(||\vec{c}_1||n) .
\end{split}
\end{align}
Using the same argument for $j$ and $k$:
\begin{align}
\begin{split}
d_1 = \iiint \limits_{\Omega} \sqrt{x^2||\vec{c}_1||^2 + y^2||\vec{c}_2||^2 + z^2||\vec{c}_3||^2}\,dx\,dy\,dz \\+ \mathcal{O}((||\vec{c}_1||+||\vec{c}_2||+||\vec{c}_3||)n(n+1)^2) .
\end{split}
\end{align}
Where $\Omega$ is a cube of parameter $n$ centered at the origin. Let's define $x' = x||\vec{c}_1||$, $y' = y||\vec{c}_2||$ and $z' = z||\vec{c}_3||$. The integral becomes:
\begin{align}
\begin{split}
d_1 = \frac{1}{V}\iiint \limits_{\Omega'} \sqrt{x'^2 + y'^2 + z'^2}dx' dy' dz' \\ + \mathcal{O}(n^3) ,
\end{split}
\end{align}
where $\Omega'$ is a prism of parameter $n||\vec{c}_1||$, $n||\vec{c}_2||$ and $n||\vec{c}_3||$ centered at the origin and $V = ||\vec{c}_1||\cdot||\vec{c}_2||\cdot||\vec{c}_3||$. This integral can be carried out in spherical coordinates by carefully adjusting the integration limits:
\begin{align}
\begin{split}
d_1 = \frac{1}{V}\sum_{i<j<k}^3\int \limits_{0}^{L_\phi} \int \limits_{0}^{L_\theta} \int \limits_{0}^{L_r}  r^3\sin{\theta}\,dr\,d\theta\,d\phi + \mathcal{O}(n^3) ,
\end{split}
\end{align}
where:
\begin{align}
L_\phi &= \tan{\left(\frac{||\vec{c}_j||}{||\vec{c}_i||}\right)} ,\\
L_\theta &= \arccos{\left(\frac{1}{\sqrt{1+\frac{||\vec{c}_i||^2}{||\vec{c}_k||^2}\sec^2{\phi}}}\right)} ,\\
L_r &= \frac{n||\vec{c}_k||}{2\cos{\theta}}.
\end{align}

Integrating:
\begin{multline}
d_1 = \frac{n^4}{192 V}\sum_{i<j<k}^3 \hspace{-1em}\int \limits_{0}^{\tan{\left(\frac{||\vec{c}_j||}{||\vec{c}_i||}\right)}} \hspace{-1em}||\vec{c}_k||^4\\((1+\frac{||\vec{c}_i||^2}{||\vec{c}_k||^2}\sec^2{\phi})^{\frac{3}{2}}-1)\,d\phi + \mathcal{O}(n^3) .
\end{multline}
Replacing $n=N^{\frac{1}{3}} - 1$, and regrouping the constants:
\begin{align}
\begin{split}
d_1 &= G_1(N^{\frac{1}{3}} - 1)^4 +  \mathcal{O}((N^{\frac{1}{3}} - 1)^3) \\
&= G_1(N^{\frac{4}{3}} - 4 N + 6 N^{\frac{2}{3}} - 4 N^{\frac{1}{3}} + 1) + \mathcal{O}(N) \\
&= G_1 N^{\frac{4}{3}} + \mathcal{O}(N) ,
\end{split}
\end{align}
where 
\begin{equation}
G_1 = \frac{1}{192 V}\sum_{i<j<k}^3 \hspace{-1em}\int \limits_{0}^{\tan{\left(\frac{||\vec{c}_j||}{||\vec{c}_i||}\right)}} \hspace{-1em}||\vec{c}_k||^4((1+\frac{||\vec{c}_i||^2}{||\vec{c}_k||^2}\sec^2{\phi})^{\frac{3}{2}}-1)\,d\phi
\end{equation}
In sum, a pure reorganization of the atoms that does not change the unit cell (equation~\eqref{eq:dist1_lin}) makes the distance depend linearly on the size whereas a pure distortion of the cell makes the distance non-linearly dependant on the size. Considering the more general case where the cell vectors are not orthonormal would only lead to lower order terms. To highlights the two leading contributions we write:
\begin{align}
d_1 = G_1N^{\frac{4}{3}} + K_1N + \mathcal{O}(N) .
\end{align}


\begin{thebibliography}{60}%
\makeatletter
\providecommand \@ifxundefined [1]{%
 \@ifx{#1\undefined}
}%
\providecommand \@ifnum [1]{%
 \ifnum #1\expandafter \@firstoftwo
 \else \expandafter \@secondoftwo
 \fi
}%
\providecommand \@ifx [1]{%
 \ifx #1\expandafter \@firstoftwo
 \else \expandafter \@secondoftwo
 \fi
}%
\providecommand \natexlab [1]{#1}%
\providecommand \enquote  [1]{``#1''}%
\providecommand \bibnamefont  [1]{#1}%
\providecommand \bibfnamefont [1]{#1}%
\providecommand \citenamefont [1]{#1}%
\providecommand \href@noop [0]{\@secondoftwo}%
\providecommand \href [0]{\begingroup \@sanitize@url \@href}%
\providecommand \@href[1]{\@@startlink{#1}\@@href}%
\providecommand \@@href[1]{\endgroup#1\@@endlink}%
\providecommand \@sanitize@url [0]{\catcode `\\12\catcode `\$12\catcode
  `\&12\catcode `\#12\catcode `\^12\catcode `\_12\catcode `\%12\relax}%
\providecommand \@@startlink[1]{}%
\providecommand \@@endlink[0]{}%
\providecommand \url  [0]{\begingroup\@sanitize@url \@url }%
\providecommand \@url [1]{\endgroup\@href {#1}{\urlprefix }}%
\providecommand \urlprefix  [0]{URL }%
\providecommand \Eprint [0]{\href }%
\providecommand \doibase [0]{http://dx.doi.org/}%
\providecommand \selectlanguage [0]{\@gobble}%
\providecommand \bibinfo  [0]{\@secondoftwo}%
\providecommand \bibfield  [0]{\@secondoftwo}%
\providecommand \translation [1]{[#1]}%
\providecommand \BibitemOpen [0]{}%
\providecommand \bibitemStop [0]{}%
\providecommand \bibitemNoStop [0]{.\EOS\space}%
\providecommand \EOS [0]{\spacefactor3000\relax}%
\providecommand \BibitemShut  [1]{\csname bibitem#1\endcsname}%
\let\auto@bib@innerbib\@empty
\bibitem [{\citenamefont {Brune}(2014)}]{brune:2014}%
  \BibitemOpen
  \bibfield  {author} {\bibinfo {author} {\bibfnamefont {H.}~\bibnamefont
  {Brune}},\ }\href@noop {} {\emph {\bibinfo {title} {Surface and Interface
  Science}}}\ (\bibinfo  {publisher} {John Wiley \& Sons, Ltd},\ \bibinfo
  {year} {2014})\ Chap.~\bibinfo {chapter} {20}, pp.\ \bibinfo {pages}
  {421--492}\BibitemShut {NoStop}%
\bibitem [{\citenamefont {Poeppelmeier}\ and\ \citenamefont
  {Rondinelli}(2016)}]{rondo_NC:2016}%
  \BibitemOpen
  \bibfield  {author} {\bibinfo {author} {\bibfnamefont {K.~R.}\ \bibnamefont
  {Poeppelmeier}}\ and\ \bibinfo {author} {\bibfnamefont {J.~M.}\ \bibnamefont
  {Rondinelli}},\ }\bibfield  {title} {\enquote {\bibinfo {title} {Mismatched
  lattices patched up},}\ }\href {https://doi.org/10.1038/nchem.2477}
  {\bibfield  {journal} {\bibinfo  {journal} {Nature Chemistry}\ }\textbf
  {\bibinfo {volume} {8}},\ \bibinfo {pages} {292 EP --} (\bibinfo {year}
  {2016})}\BibitemShut {NoStop}%
\bibitem [{\citenamefont {Ding}\ \emph {et~al.}(2016)\citenamefont {Ding},
  \citenamefont {Dwaraknath}, \citenamefont {Garten}, \citenamefont {Ndione},
  \citenamefont {Ginley},\ and\ \citenamefont {Persson}}]{persson_ACS:2016}%
  \BibitemOpen
  \bibfield  {author} {\bibinfo {author} {\bibfnamefont {H.}~\bibnamefont
  {Ding}}, \bibinfo {author} {\bibfnamefont {S.~S.}\ \bibnamefont
  {Dwaraknath}}, \bibinfo {author} {\bibfnamefont {L.}~\bibnamefont {Garten}},
  \bibinfo {author} {\bibfnamefont {P.}~\bibnamefont {Ndione}}, \bibinfo
  {author} {\bibfnamefont {D.}~\bibnamefont {Ginley}}, \ and\ \bibinfo {author}
  {\bibfnamefont {K.~A.}\ \bibnamefont {Persson}},\ }\bibfield  {title}
  {\enquote {\bibinfo {title} {Computational approach for epitaxial polymorph
  stabilization through substrate selection},}\ }\bibfield  {booktitle} {\emph
  {\bibinfo {booktitle} {ACS Applied Materials \& Interfaces}},\ }\href
  {\doibase 10.1021/acsami.6b01630} {\bibfield  {journal} {\bibinfo  {journal}
  {ACS Applied Materials \& Interfaces}\ }\textbf {\bibinfo {volume} {8}},\
  \bibinfo {pages} {13086--13093} (\bibinfo {year} {2016})}\BibitemShut
  {NoStop}%
\bibitem [{\citenamefont {Henkelman}, \citenamefont {Uberuaga},\ and\
  \citenamefont {Jónsson}(2000)}]{henkelman_2000}%
  \BibitemOpen
  \bibfield  {author} {\bibinfo {author} {\bibfnamefont {G.}~\bibnamefont
  {Henkelman}}, \bibinfo {author} {\bibfnamefont {B.~P.}\ \bibnamefont
  {Uberuaga}}, \ and\ \bibinfo {author} {\bibfnamefont {H.}~\bibnamefont
  {Jónsson}},\ }\bibfield  {title} {\enquote {\bibinfo {title} {A climbing
  image nudged elastic band method for finding saddle points and minimum energy
  paths},}\ }\href {\doibase 10.1063/1.1329672} {\bibfield  {journal} {\bibinfo
   {journal} {The Journal of Chemical Physics}\ }\textbf {\bibinfo {volume}
  {113}},\ \bibinfo {pages} {9901--9904} (\bibinfo {year} {2000})}\BibitemShut
  {NoStop}%
\bibitem [{\citenamefont {Sheppard}\ \emph {et~al.}(2012)\citenamefont
  {Sheppard}, \citenamefont {Xiao}, \citenamefont {Chemelewski}, \citenamefont
  {Johnson},\ and\ \citenamefont {Henkelman}}]{sheppard_2012}%
  \BibitemOpen
  \bibfield  {author} {\bibinfo {author} {\bibfnamefont {D.}~\bibnamefont
  {Sheppard}}, \bibinfo {author} {\bibfnamefont {P.}~\bibnamefont {Xiao}},
  \bibinfo {author} {\bibfnamefont {W.}~\bibnamefont {Chemelewski}}, \bibinfo
  {author} {\bibfnamefont {D.~D.}\ \bibnamefont {Johnson}}, \ and\ \bibinfo
  {author} {\bibfnamefont {G.}~\bibnamefont {Henkelman}},\ }\bibfield  {title}
  {\enquote {\bibinfo {title} {A generalized solid-state nudged elastic band
  method},}\ }\href@noop {} {\bibfield  {journal} {\bibinfo  {journal} {The
  Journal of chemical physics}\ }\textbf {\bibinfo {volume} {136}},\ \bibinfo
  {pages} {074103} (\bibinfo {year} {2012})}\BibitemShut {NoStop}%
\bibitem [{\citenamefont {Caspersen}\ and\ \citenamefont
  {Carter}(2005)}]{caspersen_2005}%
  \BibitemOpen
  \bibfield  {author} {\bibinfo {author} {\bibfnamefont {K.~J.}\ \bibnamefont
  {Caspersen}}\ and\ \bibinfo {author} {\bibfnamefont {E.~A.}\ \bibnamefont
  {Carter}},\ }\bibfield  {title} {\enquote {\bibinfo {title} {Finding
  transition states for crystalline solid--solid phase transformations},}\
  }\href@noop {} {\bibfield  {journal} {\bibinfo  {journal} {Proceedings of the
  National Academy of Sciences}\ }\textbf {\bibinfo {volume} {102}},\ \bibinfo
  {pages} {6738--6743} (\bibinfo {year} {2005})}\BibitemShut {NoStop}%
\bibitem [{\citenamefont {Qian}\ \emph {et~al.}(2013)\citenamefont {Qian},
  \citenamefont {Dong}, \citenamefont {Zhou}, \citenamefont {Tian},
  \citenamefont {Oganov},\ and\ \citenamefont {Wang}}]{qian_2013}%
  \BibitemOpen
  \bibfield  {author} {\bibinfo {author} {\bibfnamefont {G.-R.}\ \bibnamefont
  {Qian}}, \bibinfo {author} {\bibfnamefont {X.}~\bibnamefont {Dong}}, \bibinfo
  {author} {\bibfnamefont {X.-F.}\ \bibnamefont {Zhou}}, \bibinfo {author}
  {\bibfnamefont {Y.}~\bibnamefont {Tian}}, \bibinfo {author} {\bibfnamefont
  {A.~R.}\ \bibnamefont {Oganov}}, \ and\ \bibinfo {author} {\bibfnamefont
  {H.-T.}\ \bibnamefont {Wang}},\ }\bibfield  {title} {\enquote {\bibinfo
  {title} {Variable cell nudged elastic band method for studying solid--solid
  structural phase transitions},}\ }\href@noop {} {\bibfield  {journal}
  {\bibinfo  {journal} {Computer Physics Communications}\ }\textbf {\bibinfo
  {volume} {184}},\ \bibinfo {pages} {2111--2118} (\bibinfo {year}
  {2013})}\BibitemShut {NoStop}%
\bibitem [{Note1()}]{Note1}%
  \BibitemOpen
  \bibinfo {note} {CSL: Coincidental Site Lattice, DSC: \relax $\@@underline
  {\hbox {D}}\mathsurround \z@ $\relax isplacement of one crystal Lattice with
  respect to the second causes a pattern \relax $\@@underline {\hbox
  {S}}\mathsurround \z@ $\relax hift which is \relax $\@@underline {\hbox
  {C}}\mathsurround \z@ $\relax omplete}\BibitemShut {NoStop}%
\bibitem [{\citenamefont {Bollmann}(1970)}]{bollmann1}%
  \BibitemOpen
  \bibfield  {author} {\bibinfo {author} {\bibfnamefont {W.}~\bibnamefont
  {Bollmann}},\ }\href@noop {} {\emph {\bibinfo {title} {Crystal Defects and
  Crystalline Interfaces}}},\ \bibinfo {edition} {1st}\ ed.\ (\bibinfo
  {publisher} {Springer-Verlag Berlin Heidelberg},\ \bibinfo {year} {1970})\
  pp.\ \bibinfo {pages} {83--97}\BibitemShut {NoStop}%
\bibitem [{\citenamefont {Bollmann}(1974)}]{bollmann_1974}%
  \BibitemOpen
  \bibfield  {author} {\bibinfo {author} {\bibfnamefont {W.}~\bibnamefont
  {Bollmann}},\ }\bibfield  {title} {\enquote {\bibinfo {title} {O-lattice
  calculation of an f.c.c.-b.c.c. interface},}\ }\href {\doibase
  10.1002/pssa.2210210218} {\bibfield  {journal} {\bibinfo  {journal} {Physica
  Status Solidi (a)}\ }\textbf {\bibinfo {volume} {21}},\ \bibinfo {pages}
  {543--550} (\bibinfo {year} {1974})}\BibitemShut {NoStop}%
\bibitem [{\citenamefont {Smith}\ and\ \citenamefont
  {Pond}(1976)}]{smith_pond_1976}%
  \BibitemOpen
  \bibfield  {author} {\bibinfo {author} {\bibfnamefont {D.~A.}\ \bibnamefont
  {Smith}}\ and\ \bibinfo {author} {\bibfnamefont {R.~C.}\ \bibnamefont
  {Pond}},\ }\bibfield  {title} {\enquote {\bibinfo {title} {Bollmann's
  0-iattice theory; a geometrical approach to interface structure},}\ }\href
  {\doibase 10.1179/imtr.1976.21.1.61} {\bibfield  {journal} {\bibinfo
  {journal} {International Metals Reviews}\ }\textbf {\bibinfo {volume} {21}},\
  \bibinfo {pages} {61--74} (\bibinfo {year} {1976})},\ \Eprint
  {http://arxiv.org/abs/https://doi.org/10.1179/imtr.1976.21.1.61}
  {https://doi.org/10.1179/imtr.1976.21.1.61} \BibitemShut {NoStop}%
\bibitem [{\citenamefont {Balluffi}, \citenamefont {Brokman},\ and\
  \citenamefont {King}(1982)}]{balluffi_brokman_king_1982}%
  \BibitemOpen
  \bibfield  {author} {\bibinfo {author} {\bibfnamefont {R.}~\bibnamefont
  {Balluffi}}, \bibinfo {author} {\bibfnamefont {A.}~\bibnamefont {Brokman}}, \
  and\ \bibinfo {author} {\bibfnamefont {A.}~\bibnamefont {King}},\ }\bibfield
  {title} {\enquote {\bibinfo {title} {Csl/dsc lattice model for general
  crystalcrystal boundaries and their line defects},}\ }\href {\doibase
  10.1016/0001-6160(82)90166-3} {\bibfield  {journal} {\bibinfo  {journal}
  {Acta Metallurgica}\ }\textbf {\bibinfo {volume} {30}},\ \bibinfo {pages}
  {1453--1470} (\bibinfo {year} {1982})}\BibitemShut {NoStop}%
\bibitem [{\citenamefont {Zhang}\ and\ \citenamefont
  {Kelly}(1998)}]{zhang_kelly_1998}%
  \BibitemOpen
  \bibfield  {author} {\bibinfo {author} {\bibfnamefont {M.-X.}\ \bibnamefont
  {Zhang}}\ and\ \bibinfo {author} {\bibfnamefont {P.}~\bibnamefont {Kelly}},\
  }\bibfield  {title} {\enquote {\bibinfo {title} {Crystallography and
  morphology of widmanstätten cementite in austenite},}\ }\href {\doibase
  10.1016/s1359-6454(98)00139-6} {\bibfield  {journal} {\bibinfo  {journal}
  {Acta Materialia}\ }\textbf {\bibinfo {volume} {46}},\ \bibinfo {pages}
  {4617--4628} (\bibinfo {year} {1998})}\BibitemShut {NoStop}%
\bibitem [{\citenamefont {Zhang}\ \emph {et~al.}(2005)\citenamefont {Zhang},
  \citenamefont {Kelly}, \citenamefont {Easton},\ and\ \citenamefont
  {Taylor}}]{zhang_kelly_easton_taylor_2005}%
  \BibitemOpen
  \bibfield  {author} {\bibinfo {author} {\bibfnamefont {M.}~\bibnamefont
  {Zhang}}, \bibinfo {author} {\bibfnamefont {P.}~\bibnamefont {Kelly}},
  \bibinfo {author} {\bibfnamefont {M.}~\bibnamefont {Easton}}, \ and\ \bibinfo
  {author} {\bibfnamefont {J.}~\bibnamefont {Taylor}},\ }\bibfield  {title}
  {\enquote {\bibinfo {title} {Crystallographic study of grain refinement in
  aluminum alloys using the edge-to-edge matching model},}\ }\href {\doibase
  10.1016/j.actamat.2004.11.037} {\bibfield  {journal} {\bibinfo  {journal}
  {Acta Materialia}\ }\textbf {\bibinfo {volume} {53}},\ \bibinfo {pages}
  {1427--1438} (\bibinfo {year} {2005})}\BibitemShut {NoStop}%
\bibitem [{\citenamefont {Ikuhara}\ and\ \citenamefont
  {Pirouz}(1996)}]{ikuhara_pirouz_1996}%
  \BibitemOpen
  \bibfield  {author} {\bibinfo {author} {\bibfnamefont {Y.}~\bibnamefont
  {Ikuhara}}\ and\ \bibinfo {author} {\bibfnamefont {P.}~\bibnamefont
  {Pirouz}},\ }\bibfield  {title} {\enquote {\bibinfo {title} {Orientation
  relationship in large mismatched bicrystals and coincidence of reciprocal
  lattice points (crlp)},}\ }\href {\doibase
  10.4028/www.scientific.net/msf.207-209.121} {\bibfield  {journal} {\bibinfo
  {journal} {Materials Science Forum}\ }\textbf {\bibinfo {volume} {207-209}},\
  \bibinfo {pages} {121--124} (\bibinfo {year} {1996})}\BibitemShut {NoStop}%
\bibitem [{\citenamefont {Zur}\ and\ \citenamefont {McGill}(1984)}]{zur1984}%
  \BibitemOpen
  \bibfield  {author} {\bibinfo {author} {\bibfnamefont {A.}~\bibnamefont
  {Zur}}\ and\ \bibinfo {author} {\bibfnamefont {T.~C.}\ \bibnamefont
  {McGill}},\ }\bibfield  {title} {\enquote {\bibinfo {title} {Lattice match:
  An application to heteroepitaxy},}\ }\href {\doibase 10.1063/1.333084}
  {\bibfield  {journal} {\bibinfo  {journal} {Journal of Applied Physics}\
  }\textbf {\bibinfo {volume} {55}},\ \bibinfo {pages} {378--386} (\bibinfo
  {year} {1984})}\BibitemShut {NoStop}%
\bibitem [{\citenamefont {Mathew}\ \emph {et~al.}(2016)\citenamefont {Mathew},
  \citenamefont {Singh}, \citenamefont {Gabriel}, \citenamefont {Choudhary},
  \citenamefont {Sinnott}, \citenamefont {Davydov}, \citenamefont {Tavazza},\
  and\ \citenamefont {Hennig}}]{mathew_2016}%
  \BibitemOpen
  \bibfield  {author} {\bibinfo {author} {\bibfnamefont {K.}~\bibnamefont
  {Mathew}}, \bibinfo {author} {\bibfnamefont {A.~K.}\ \bibnamefont {Singh}},
  \bibinfo {author} {\bibfnamefont {J.~J.}\ \bibnamefont {Gabriel}}, \bibinfo
  {author} {\bibfnamefont {K.}~\bibnamefont {Choudhary}}, \bibinfo {author}
  {\bibfnamefont {S.~B.}\ \bibnamefont {Sinnott}}, \bibinfo {author}
  {\bibfnamefont {A.~V.}\ \bibnamefont {Davydov}}, \bibinfo {author}
  {\bibfnamefont {F.}~\bibnamefont {Tavazza}}, \ and\ \bibinfo {author}
  {\bibfnamefont {R.~G.}\ \bibnamefont {Hennig}},\ }\bibfield  {title}
  {\enquote {\bibinfo {title} {Mpinterfaces: A materials project based python
  tool for high-throughput computational screening of interfacial systems},}\
  }\href@noop {} {\bibfield  {journal} {\bibinfo  {journal} {Computational
  Materials Science}\ }\textbf {\bibinfo {volume} {122}},\ \bibinfo {pages}
  {183--190} (\bibinfo {year} {2016})}\BibitemShut {NoStop}%
\bibitem [{\citenamefont {Jelver}\ \emph {et~al.}(2017)\citenamefont {Jelver},
  \citenamefont {Larsen}, \citenamefont {Stradi}, \citenamefont {Stokbro},\
  and\ \citenamefont {Jacobsen}}]{jelver2017determination}%
  \BibitemOpen
  \bibfield  {author} {\bibinfo {author} {\bibfnamefont {L.}~\bibnamefont
  {Jelver}}, \bibinfo {author} {\bibfnamefont {P.~M.}\ \bibnamefont {Larsen}},
  \bibinfo {author} {\bibfnamefont {D.}~\bibnamefont {Stradi}}, \bibinfo
  {author} {\bibfnamefont {K.}~\bibnamefont {Stokbro}}, \ and\ \bibinfo
  {author} {\bibfnamefont {K.~W.}\ \bibnamefont {Jacobsen}},\ }\bibfield
  {title} {\enquote {\bibinfo {title} {Determination of low-strain interfaces
  via geometric matching},}\ }\href@noop {} {\bibfield  {journal} {\bibinfo
  {journal} {Physical Review B}\ }\textbf {\bibinfo {volume} {96}},\ \bibinfo
  {pages} {085306} (\bibinfo {year} {2017})}\BibitemShut {NoStop}%
\bibitem [{\citenamefont {Sadeghi}\ \emph {et~al.}(2013)\citenamefont
  {Sadeghi}, \citenamefont {Ghasemi}, \citenamefont {Schaefer}, \citenamefont
  {Mohr}, \citenamefont {Lill},\ and\ \citenamefont
  {Goedecker}}]{sadeghi_2013}%
  \BibitemOpen
  \bibfield  {author} {\bibinfo {author} {\bibfnamefont {A.}~\bibnamefont
  {Sadeghi}}, \bibinfo {author} {\bibfnamefont {S.~A.}\ \bibnamefont
  {Ghasemi}}, \bibinfo {author} {\bibfnamefont {B.}~\bibnamefont {Schaefer}},
  \bibinfo {author} {\bibfnamefont {S.}~\bibnamefont {Mohr}}, \bibinfo {author}
  {\bibfnamefont {M.~A.}\ \bibnamefont {Lill}}, \ and\ \bibinfo {author}
  {\bibfnamefont {S.}~\bibnamefont {Goedecker}},\ }\bibfield  {title} {\enquote
  {\bibinfo {title} {Metrics for measuring distances in configuration
  spaces},}\ }\href@noop {} {\bibfield  {journal} {\bibinfo  {journal} {The
  Journal of chemical physics}\ }\textbf {\bibinfo {volume} {139}},\ \bibinfo
  {pages} {184118} (\bibinfo {year} {2013})}\BibitemShut {NoStop}%
\bibitem [{Note2()}]{Note2}%
  \BibitemOpen
  \bibinfo {note} {The $l_2$-norm in configuration space equivalent to the
  Forbenius norm of the position matrix}\BibitemShut {NoStop}%
\bibitem [{\citenamefont {Oganov}\ and\ \citenamefont
  {Valle}(2009)}]{oganov_JCP:2009}%
  \BibitemOpen
  \bibfield  {author} {\bibinfo {author} {\bibfnamefont {A.~R.}\ \bibnamefont
  {Oganov}}\ and\ \bibinfo {author} {\bibfnamefont {M.}~\bibnamefont {Valle}},\
  }\bibfield  {title} {\enquote {\bibinfo {title} {How to quantify energy
  landscapes of solids},}\ }\href {\doibase 10.1063/1.3079326} {\bibfield
  {journal} {\bibinfo  {journal} {The Journal of Chemical Physics}\ }\textbf
  {\bibinfo {volume} {130}},\ \bibinfo {pages} {104504} (\bibinfo {year}
  {2009})}\BibitemShut {NoStop}%
\bibitem [{\citenamefont {Bart{\'o}k}, \citenamefont {Kondor},\ and\
  \citenamefont {Cs{\'a}nyi}(2013)}]{bartoK_2013}%
  \BibitemOpen
  \bibfield  {author} {\bibinfo {author} {\bibfnamefont {A.~P.}\ \bibnamefont
  {Bart{\'o}k}}, \bibinfo {author} {\bibfnamefont {R.}~\bibnamefont {Kondor}},
  \ and\ \bibinfo {author} {\bibfnamefont {G.}~\bibnamefont {Cs{\'a}nyi}},\
  }\bibfield  {title} {\enquote {\bibinfo {title} {On representing chemical
  environments},}\ }\href@noop {} {\bibfield  {journal} {\bibinfo  {journal}
  {Physical Review B}\ }\textbf {\bibinfo {volume} {87}},\ \bibinfo {pages}
  {184115} (\bibinfo {year} {2013})}\BibitemShut {NoStop}%
\bibitem [{\citenamefont {Yang}, \citenamefont {Dacek},\ and\ \citenamefont
  {Ceder}(2014)}]{yang_2014}%
  \BibitemOpen
  \bibfield  {author} {\bibinfo {author} {\bibfnamefont {L.}~\bibnamefont
  {Yang}}, \bibinfo {author} {\bibfnamefont {S.}~\bibnamefont {Dacek}}, \ and\
  \bibinfo {author} {\bibfnamefont {G.}~\bibnamefont {Ceder}},\ }\bibfield
  {title} {\enquote {\bibinfo {title} {Proposed definition of crystal
  substructure and substructural similarity},}\ }\href@noop {} {\bibfield
  {journal} {\bibinfo  {journal} {Physical Review B}\ }\textbf {\bibinfo
  {volume} {90}},\ \bibinfo {pages} {054102} (\bibinfo {year}
  {2014})}\BibitemShut {NoStop}%
\bibitem [{\citenamefont {De}\ \emph {et~al.}(2016)\citenamefont {De},
  \citenamefont {Bart{\'o}k}, \citenamefont {Cs{\'a}nyi},\ and\ \citenamefont
  {Ceriotti}}]{de_2016}%
  \BibitemOpen
  \bibfield  {author} {\bibinfo {author} {\bibfnamefont {S.}~\bibnamefont
  {De}}, \bibinfo {author} {\bibfnamefont {A.~P.}\ \bibnamefont {Bart{\'o}k}},
  \bibinfo {author} {\bibfnamefont {G.}~\bibnamefont {Cs{\'a}nyi}}, \ and\
  \bibinfo {author} {\bibfnamefont {M.}~\bibnamefont {Ceriotti}},\ }\bibfield
  {title} {\enquote {\bibinfo {title} {Comparing molecules and solids across
  structural and alchemical space},}\ }\href@noop {} {\bibfield  {journal}
  {\bibinfo  {journal} {Physical Chemistry Chemical Physics}\ }\textbf
  {\bibinfo {volume} {18}},\ \bibinfo {pages} {13754--13769} (\bibinfo {year}
  {2016})}\BibitemShut {NoStop}%
\bibitem [{\citenamefont {Zhu}\ \emph {et~al.}(2016)\citenamefont {Zhu},
  \citenamefont {Amsler}, \citenamefont {Fuhrer}, \citenamefont {Schaefer},
  \citenamefont {Faraji}, \citenamefont {Rostami}, \citenamefont {Ghasemi},
  \citenamefont {Sadeghi}, \citenamefont {Grauzinyte}, \citenamefont
  {Wolverton} \emph {et~al.}}]{zhu_2016}%
  \BibitemOpen
  \bibfield  {author} {\bibinfo {author} {\bibfnamefont {L.}~\bibnamefont
  {Zhu}}, \bibinfo {author} {\bibfnamefont {M.}~\bibnamefont {Amsler}},
  \bibinfo {author} {\bibfnamefont {T.}~\bibnamefont {Fuhrer}}, \bibinfo
  {author} {\bibfnamefont {B.}~\bibnamefont {Schaefer}}, \bibinfo {author}
  {\bibfnamefont {S.}~\bibnamefont {Faraji}}, \bibinfo {author} {\bibfnamefont
  {S.}~\bibnamefont {Rostami}}, \bibinfo {author} {\bibfnamefont {S.~A.}\
  \bibnamefont {Ghasemi}}, \bibinfo {author} {\bibfnamefont {A.}~\bibnamefont
  {Sadeghi}}, \bibinfo {author} {\bibfnamefont {M.}~\bibnamefont {Grauzinyte}},
  \bibinfo {author} {\bibfnamefont {C.}~\bibnamefont {Wolverton}},  \emph
  {et~al.},\ }\bibfield  {title} {\enquote {\bibinfo {title} {A fingerprint
  based metric for measuring similarities of crystalline structures},}\
  }\href@noop {} {\bibfield  {journal} {\bibinfo  {journal} {The Journal of
  chemical physics}\ }\textbf {\bibinfo {volume} {144}},\ \bibinfo {pages}
  {034203} (\bibinfo {year} {2016})}\BibitemShut {NoStop}%
\bibitem [{\citenamefont {Niggli}(1928)}]{niggli1928handbuch}%
  \BibitemOpen
  \bibfield  {author} {\bibinfo {author} {\bibfnamefont {P.}~\bibnamefont
  {Niggli}},\ }\href@noop {} {\emph {\bibinfo {title} {Handbuch der
  Experimentalphysik}}},\ Vol.~\bibinfo {volume} {7}\ (\bibinfo  {publisher}
  {Akademische Verlagsgesellschaft},\ \bibinfo {year} {1928})\ \bibinfo {note}
  {part 1}\BibitemShut {NoStop}%
\bibitem [{\citenamefont {Santoro}\ and\ \citenamefont
  {Mighell}(1970)}]{santoro1970determination}%
  \BibitemOpen
  \bibfield  {author} {\bibinfo {author} {\bibfnamefont {A.~t.}\ \bibnamefont
  {Santoro}}\ and\ \bibinfo {author} {\bibfnamefont {A.}~\bibnamefont
  {Mighell}},\ }\bibfield  {title} {\enquote {\bibinfo {title} {Determination
  of reduced cells},}\ }\href@noop {} {\bibfield  {journal} {\bibinfo
  {journal} {Acta Crystallographica Section A: Crystal Physics, Diffraction,
  Theoretical and General Crystallography}\ }\textbf {\bibinfo {volume} {26}},\
  \bibinfo {pages} {124--127} (\bibinfo {year} {1970})}\BibitemShut {NoStop}%
\bibitem [{\citenamefont {Kriv\'y}\ and\ \citenamefont
  {Gruber}(1976)}]{krivy1976}%
  \BibitemOpen
  \bibfield  {author} {\bibinfo {author} {\bibfnamefont {I.}~\bibnamefont
  {Kriv\'y}}\ and\ \bibinfo {author} {\bibfnamefont {B.}~\bibnamefont
  {Gruber}},\ }\bibfield  {title} {\enquote {\bibinfo {title} {A unified
  algorithm for determining the reduced (niggli) cell},}\ }\href {\doibase
  10.1107/s0567739476000636} {\bibfield  {journal} {\bibinfo  {journal} {Acta
  Crystallographica Section A}\ }\textbf {\bibinfo {volume} {32}},\ \bibinfo
  {pages} {297--298} (\bibinfo {year} {1976})}\BibitemShut {NoStop}%
\bibitem [{\citenamefont {Stevanovi\'c}\ \emph {et~al.}(2018)\citenamefont
  {Stevanovi\'c}, \citenamefont {Trottier}, \citenamefont {Musgrave},
  \citenamefont {Therrien}, \citenamefont {Holder},\ and\ \citenamefont
  {Graf}}]{stevanovic2018}%
  \BibitemOpen
  \bibfield  {author} {\bibinfo {author} {\bibfnamefont {V.}~\bibnamefont
  {Stevanovi\'c}}, \bibinfo {author} {\bibfnamefont {R.}~\bibnamefont
  {Trottier}}, \bibinfo {author} {\bibfnamefont {C.}~\bibnamefont {Musgrave}},
  \bibinfo {author} {\bibfnamefont {F.}~\bibnamefont {Therrien}}, \bibinfo
  {author} {\bibfnamefont {A.}~\bibnamefont {Holder}}, \ and\ \bibinfo {author}
  {\bibfnamefont {P.}~\bibnamefont {Graf}},\ }\bibfield  {title} {\enquote
  {\bibinfo {title} {Predicting kinetics of polymorphic transformations from
  structure mapping and coordination analysis},}\ }\href {\doibase
  10.1103/physrevmaterials.2.033802} {\bibfield  {journal} {\bibinfo  {journal}
  {Physical Review Materials}\ }\textbf {\bibinfo {volume} {2}} (\bibinfo
  {year} {2018}),\ 10.1103/physrevmaterials.2.033802}\BibitemShut {NoStop}%
\bibitem [{\citenamefont {Lonie}\ and\ \citenamefont
  {Zurek}(2012)}]{lonie_2012}%
  \BibitemOpen
  \bibfield  {author} {\bibinfo {author} {\bibfnamefont {D.~C.}\ \bibnamefont
  {Lonie}}\ and\ \bibinfo {author} {\bibfnamefont {E.}~\bibnamefont {Zurek}},\
  }\bibfield  {title} {\enquote {\bibinfo {title} {Identifying duplicate
  crystal structures: Xtalcomp, an open-source solution},}\ }\href@noop {}
  {\bibfield  {journal} {\bibinfo  {journal} {Computer Physics Communications}\
  }\textbf {\bibinfo {volume} {183}},\ \bibinfo {pages} {690--697} (\bibinfo
  {year} {2012})}\BibitemShut {NoStop}%
\bibitem [{\citenamefont {Capillas}, \citenamefont {Perez-Mato},\ and\
  \citenamefont {Aroyo}(2007)}]{capillas_2007}%
  \BibitemOpen
  \bibfield  {author} {\bibinfo {author} {\bibfnamefont {C.}~\bibnamefont
  {Capillas}}, \bibinfo {author} {\bibfnamefont {J.}~\bibnamefont
  {Perez-Mato}}, \ and\ \bibinfo {author} {\bibfnamefont {M.}~\bibnamefont
  {Aroyo}},\ }\bibfield  {title} {\enquote {\bibinfo {title} {Maximal symmetry
  transition paths for reconstructive phase transitions},}\ }\href@noop {}
  {\bibfield  {journal} {\bibinfo  {journal} {Journal of Physics: Condensed
  Matter}\ }\textbf {\bibinfo {volume} {19}},\ \bibinfo {pages} {275203}
  (\bibinfo {year} {2007})}\BibitemShut {NoStop}%
\bibitem [{\citenamefont {Larsen}, \citenamefont {Schi{\o}tz},\ and\
  \citenamefont {Schmidt}(2017)}]{larsen2017structural}%
  \BibitemOpen
  \bibfield  {author} {\bibinfo {author} {\bibfnamefont {P.~M.}\ \bibnamefont
  {Larsen}}, \bibinfo {author} {\bibfnamefont {J.}~\bibnamefont {Schi{\o}tz}},
  \ and\ \bibinfo {author} {\bibfnamefont {S.}~\bibnamefont {Schmidt}},\
  }\href@noop {} {\emph {\bibinfo {title} {Structural analysis algorithms for
  nanomaterials}}}\ (\bibinfo  {publisher} {Department of Physics, Technical
  University of Denmark},\ \bibinfo {year} {2017})\BibitemShut {NoStop}%
\bibitem [{\citenamefont {Zhu}\ \emph {et~al.}(2019)\citenamefont {Zhu},
  \citenamefont {Guo}, \citenamefont {Zou}, \citenamefont {Li}, \citenamefont
  {Yuen}, \citenamefont {Mihaylova},\ and\ \citenamefont
  {Leung}}]{zhu2019review}%
  \BibitemOpen
  \bibfield  {author} {\bibinfo {author} {\bibfnamefont {H.}~\bibnamefont
  {Zhu}}, \bibinfo {author} {\bibfnamefont {B.}~\bibnamefont {Guo}}, \bibinfo
  {author} {\bibfnamefont {K.}~\bibnamefont {Zou}}, \bibinfo {author}
  {\bibfnamefont {Y.}~\bibnamefont {Li}}, \bibinfo {author} {\bibfnamefont
  {K.-V.}\ \bibnamefont {Yuen}}, \bibinfo {author} {\bibfnamefont
  {L.}~\bibnamefont {Mihaylova}}, \ and\ \bibinfo {author} {\bibfnamefont
  {H.}~\bibnamefont {Leung}},\ }\bibfield  {title} {\enquote {\bibinfo {title}
  {A review of point set registration: From pairwise registration to groupwise
  registration},}\ }\href@noop {} {\bibfield  {journal} {\bibinfo  {journal}
  {Sensors}\ }\textbf {\bibinfo {volume} {19}},\ \bibinfo {pages} {1191}
  (\bibinfo {year} {2019})}\BibitemShut {NoStop}%
\bibitem [{\citenamefont {Besl}\ and\ \citenamefont {McKay}(1992)}]{besl_1992}%
  \BibitemOpen
  \bibfield  {author} {\bibinfo {author} {\bibfnamefont {P.~J.}\ \bibnamefont
  {Besl}}\ and\ \bibinfo {author} {\bibfnamefont {N.~D.}\ \bibnamefont
  {McKay}},\ }\bibfield  {title} {\enquote {\bibinfo {title} {{Method for
  registration of 3-D shapes}},}\ }in\ \href {\doibase 10.1117/12.57955} {\emph
  {\bibinfo {booktitle} {Sensor Fusion IV: Control Paradigms and Data
  Structures}}},\ Vol.\ \bibinfo {volume} {1611},\ \bibinfo {editor} {edited
  by\ \bibinfo {editor} {\bibfnamefont {P.~S.}\ \bibnamefont {Schenker}}},\
  \bibinfo {organization} {International Society for Optics and Photonics}\
  (\bibinfo  {publisher} {SPIE},\ \bibinfo {year} {1992})\ pp.\ \bibinfo
  {pages} {586 -- 606}\BibitemShut {NoStop}%
\bibitem [{\citenamefont {Pan}\ \emph {et~al.}(2018)\citenamefont {Pan},
  \citenamefont {Yang}, \citenamefont {Liang},\ and\ \citenamefont
  {Dong}}]{pan2018iterative}%
  \BibitemOpen
  \bibfield  {author} {\bibinfo {author} {\bibfnamefont {Y.}~\bibnamefont
  {Pan}}, \bibinfo {author} {\bibfnamefont {B.}~\bibnamefont {Yang}}, \bibinfo
  {author} {\bibfnamefont {F.}~\bibnamefont {Liang}}, \ and\ \bibinfo {author}
  {\bibfnamefont {Z.}~\bibnamefont {Dong}},\ }\bibfield  {title} {\enquote
  {\bibinfo {title} {Iterative global similarity points: A robust
  coarse-to-fine integration solution for pairwise 3d point cloud
  registration},}\ }in\ \href@noop {} {\emph {\bibinfo {booktitle} {2018
  International Conference on 3D Vision (3DV)}}}\ (\bibinfo {organization}
  {IEEE},\ \bibinfo {year} {2018})\ pp.\ \bibinfo {pages}
  {180--189}\BibitemShut {NoStop}%
\bibitem [{\citenamefont {Bhandarkar}\ \emph {et~al.}(2004)\citenamefont
  {Bhandarkar}, \citenamefont {Chowdhury}, \citenamefont {Tang}, \citenamefont
  {Yu},\ and\ \citenamefont {Tolllner}}]{bhandarkar2004surface}%
  \BibitemOpen
  \bibfield  {author} {\bibinfo {author} {\bibfnamefont {S.~M.}\ \bibnamefont
  {Bhandarkar}}, \bibinfo {author} {\bibfnamefont {A.~S.}\ \bibnamefont
  {Chowdhury}}, \bibinfo {author} {\bibfnamefont {Y.}~\bibnamefont {Tang}},
  \bibinfo {author} {\bibfnamefont {J.}~\bibnamefont {Yu}}, \ and\ \bibinfo
  {author} {\bibfnamefont {E.}~\bibnamefont {Tolllner}},\ }\bibfield  {title}
  {\enquote {\bibinfo {title} {Surface matching algorithms computer aided
  reconstructive plastic surgery},}\ }in\ \href@noop {} {\emph {\bibinfo
  {booktitle} {2004 2nd IEEE International Symposium on Biomedical Imaging:
  Nano to Macro (IEEE Cat No. 04EX821)}}}\ (\bibinfo {organization} {IEEE},\
  \bibinfo {year} {2004})\ pp.\ \bibinfo {pages} {740--743}\BibitemShut
  {NoStop}%
\bibitem [{\citenamefont {Maiseli}, \citenamefont {Gu},\ and\ \citenamefont
  {Gao}(2017)}]{maiseli2017recent}%
  \BibitemOpen
  \bibfield  {author} {\bibinfo {author} {\bibfnamefont {B.}~\bibnamefont
  {Maiseli}}, \bibinfo {author} {\bibfnamefont {Y.}~\bibnamefont {Gu}}, \ and\
  \bibinfo {author} {\bibfnamefont {H.}~\bibnamefont {Gao}},\ }\bibfield
  {title} {\enquote {\bibinfo {title} {Recent developments and trends in point
  set registration methods},}\ }\href@noop {} {\bibfield  {journal} {\bibinfo
  {journal} {Journal of Visual Communication and Image Representation}\
  }\textbf {\bibinfo {volume} {46}},\ \bibinfo {pages} {95--106} (\bibinfo
  {year} {2017})}\BibitemShut {NoStop}%
\bibitem [{\citenamefont {Kuhn}(1955)}]{kuhn1955}%
  \BibitemOpen
  \bibfield  {author} {\bibinfo {author} {\bibfnamefont {H.~W.}\ \bibnamefont
  {Kuhn}},\ }\bibfield  {title} {\enquote {\bibinfo {title} {The hungarian
  method for the assignment problem},}\ }\href {\doibase
  10.1002/nav.3800020109} {\bibfield  {journal} {\bibinfo  {journal} {Naval
  Research Logistics Quarterly}\ }\textbf {\bibinfo {volume} {2}},\ \bibinfo
  {pages} {83--97} (\bibinfo {year} {1955})}\BibitemShut {NoStop}%
\bibitem [{\citenamefont {Burgers}(1934)}]{burgers_1934}%
  \BibitemOpen
  \bibfield  {author} {\bibinfo {author} {\bibfnamefont {W.}~\bibnamefont
  {Burgers}},\ }\bibfield  {title} {\enquote {\bibinfo {title} {On the process
  of transition of the cubic-body-centered modification into the
  hexagonal-close-packed modification of zirconium},}\ }\href@noop {}
  {\bibfield  {journal} {\bibinfo  {journal} {Physica}\ }\textbf {\bibinfo
  {volume} {1}},\ \bibinfo {pages} {561--586} (\bibinfo {year}
  {1934})}\BibitemShut {NoStop}%
\bibitem [{\citenamefont {Masuda-Jindo}, \citenamefont {Nishitani},\ and\
  \citenamefont {Van~Hung}(2004)}]{masuda_2004}%
  \BibitemOpen
  \bibfield  {author} {\bibinfo {author} {\bibfnamefont {K.}~\bibnamefont
  {Masuda-Jindo}}, \bibinfo {author} {\bibfnamefont {S.}~\bibnamefont
  {Nishitani}}, \ and\ \bibinfo {author} {\bibfnamefont {V.}~\bibnamefont
  {Van~Hung}},\ }\bibfield  {title} {\enquote {\bibinfo {title} {hcp-bcc
  structural phase transformation of titanium: Analytic model calculations},}\
  }\href@noop {} {\bibfield  {journal} {\bibinfo  {journal} {Physical Review
  B}\ }\textbf {\bibinfo {volume} {70}},\ \bibinfo {pages} {184122} (\bibinfo
  {year} {2004})}\BibitemShut {NoStop}%
\bibitem [{\citenamefont {Khaliullin}\ \emph {et~al.}(2011)\citenamefont
  {Khaliullin}, \citenamefont {Eshet}, \citenamefont {K{\"u}hne}, \citenamefont
  {Behler},\ and\ \citenamefont {Parrinello}}]{khaliullin_2011}%
  \BibitemOpen
  \bibfield  {author} {\bibinfo {author} {\bibfnamefont {R.~Z.}\ \bibnamefont
  {Khaliullin}}, \bibinfo {author} {\bibfnamefont {H.}~\bibnamefont {Eshet}},
  \bibinfo {author} {\bibfnamefont {T.~D.}\ \bibnamefont {K{\"u}hne}}, \bibinfo
  {author} {\bibfnamefont {J.}~\bibnamefont {Behler}}, \ and\ \bibinfo {author}
  {\bibfnamefont {M.}~\bibnamefont {Parrinello}},\ }\bibfield  {title}
  {\enquote {\bibinfo {title} {Nucleation mechanism for the direct
  graphite-to-diamond phase transition},}\ }\href@noop {} {\bibfield  {journal}
  {\bibinfo  {journal} {Nature materials}\ }\textbf {\bibinfo {volume} {10}},\
  \bibinfo {pages} {693} (\bibinfo {year} {2011})}\BibitemShut {NoStop}%
\bibitem [{\citenamefont {Xiao}\ and\ \citenamefont
  {Henkelman}(2012)}]{xiao_2012}%
  \BibitemOpen
  \bibfield  {author} {\bibinfo {author} {\bibfnamefont {P.}~\bibnamefont
  {Xiao}}\ and\ \bibinfo {author} {\bibfnamefont {G.}~\bibnamefont
  {Henkelman}},\ }\bibfield  {title} {\enquote {\bibinfo {title}
  {Communication: From graphite to diamond: Reaction pathways of the phase
  transition},}\ }\href {\doibase 10.1063/1.4752249} {\bibfield  {journal}
  {\bibinfo  {journal} {The Journal of Chemical Physics}\ }\textbf {\bibinfo
  {volume} {137}},\ \bibinfo {pages} {101101} (\bibinfo {year}
  {2012})}\BibitemShut {NoStop}%
\bibitem [{\citenamefont {Bain}\ and\ \citenamefont
  {Dunkirk}(1924)}]{bain_1924}%
  \BibitemOpen
  \bibfield  {author} {\bibinfo {author} {\bibfnamefont {E.~C.}\ \bibnamefont
  {Bain}}\ and\ \bibinfo {author} {\bibfnamefont {N.}~\bibnamefont {Dunkirk}},\
  }\bibfield  {title} {\enquote {\bibinfo {title} {The nature of martensite},}\
  }\href@noop {} {\bibfield  {journal} {\bibinfo  {journal} {trans. AIME}\
  }\textbf {\bibinfo {volume} {70}},\ \bibinfo {pages} {25--47} (\bibinfo
  {year} {1924})}\BibitemShut {NoStop}%
\bibitem [{\citenamefont {Nishiyama}(2012)}]{nishiyama_2012}%
  \BibitemOpen
  \bibfield  {author} {\bibinfo {author} {\bibfnamefont {Z.}~\bibnamefont
  {Nishiyama}},\ }\href@noop {} {\emph {\bibinfo {title} {Martensitic
  transformation}}}\ (\bibinfo  {publisher} {Elsevier},\ \bibinfo {year}
  {2012})\BibitemShut {NoStop}%
\bibitem [{\citenamefont {Watanabe}, \citenamefont {Tokonami},\ and\
  \citenamefont {Morimoto}(1977)}]{watanabe_1977}%
  \BibitemOpen
  \bibfield  {author} {\bibinfo {author} {\bibfnamefont {M.}~\bibnamefont
  {Watanabe}}, \bibinfo {author} {\bibfnamefont {M.}~\bibnamefont {Tokonami}},
  \ and\ \bibinfo {author} {\bibfnamefont {N.}~\bibnamefont {Morimoto}},\
  }\bibfield  {title} {\enquote {\bibinfo {title} {The transition mechanism
  between the cscl-type and nacl-type structures in cscl},}\ }\href@noop {}
  {\bibfield  {journal} {\bibinfo  {journal} {Acta Crystallographica Section A:
  Crystal Physics, Diffraction, Theoretical and General Crystallography}\
  }\textbf {\bibinfo {volume} {33}},\ \bibinfo {pages} {294--298} (\bibinfo
  {year} {1977})}\BibitemShut {NoStop}%
\bibitem [{\citenamefont {Blanco}\ \emph {et~al.}(2000)\citenamefont {Blanco},
  \citenamefont {Recio}, \citenamefont {Costales},\ and\ \citenamefont
  {Pandey}}]{blanco_2000}%
  \BibitemOpen
  \bibfield  {author} {\bibinfo {author} {\bibfnamefont {M.~A.}\ \bibnamefont
  {Blanco}}, \bibinfo {author} {\bibfnamefont {J.~M.}\ \bibnamefont {Recio}},
  \bibinfo {author} {\bibfnamefont {A.}~\bibnamefont {Costales}}, \ and\
  \bibinfo {author} {\bibfnamefont {R.}~\bibnamefont {Pandey}},\ }\bibfield
  {title} {\enquote {\bibinfo {title} {Transition path for the
  $b3\ensuremath{\rightleftharpoons}b1$ phase transformation in
  semiconductors},}\ }\href {\doibase 10.1103/PhysRevB.62.R10599} {\bibfield
  {journal} {\bibinfo  {journal} {Phys. Rev. B}\ }\textbf {\bibinfo {volume}
  {62}},\ \bibinfo {pages} {R10599--R10602} (\bibinfo {year}
  {2000})}\BibitemShut {NoStop}%
\bibitem [{\citenamefont {Sowa}(2001)}]{sowa_2001}%
  \BibitemOpen
  \bibfield  {author} {\bibinfo {author} {\bibfnamefont {H.}~\bibnamefont
  {Sowa}},\ }\bibfield  {title} {\enquote {\bibinfo {title} {On the transition
  from the wurtzite to the nacl type},}\ }\href@noop {} {\bibfield  {journal}
  {\bibinfo  {journal} {Acta Crystallographica Section A: Foundations of
  Crystallography}\ }\textbf {\bibinfo {volume} {57}},\ \bibinfo {pages}
  {176--182} (\bibinfo {year} {2001})}\BibitemShut {NoStop}%
\bibitem [{\citenamefont {Zhang}\ and\ \citenamefont
  {Kelly}(2005)}]{zhang_kelly_2005}%
  \BibitemOpen
  \bibfield  {author} {\bibinfo {author} {\bibfnamefont {M.-X.}\ \bibnamefont
  {Zhang}}\ and\ \bibinfo {author} {\bibfnamefont {P.}~\bibnamefont {Kelly}},\
  }\bibfield  {title} {\enquote {\bibinfo {title} {Edge-to-edge matching model
  for predicting orientation relationships and habit planes—the
  improvements},}\ }\href {\doibase
  https://doi.org/10.1016/j.scriptamat.2005.01.040} {\bibfield  {journal}
  {\bibinfo  {journal} {Scripta Materialia}\ }\textbf {\bibinfo {volume}
  {52}},\ \bibinfo {pages} {963 -- 968} (\bibinfo {year} {2005})}\BibitemShut
  {NoStop}%
\bibitem [{\citenamefont {Bowles}(1951)}]{bowles1951crystallographic}%
  \BibitemOpen
  \bibfield  {author} {\bibinfo {author} {\bibfnamefont {J.}~\bibnamefont
  {Bowles}},\ }\bibfield  {title} {\enquote {\bibinfo {title} {The
  crystallographic mechanism of the martensite reaction in iron-carbon
  alloys},}\ }\href@noop {} {\bibfield  {journal} {\bibinfo  {journal} {Acta
  Crystallographica}\ }\textbf {\bibinfo {volume} {4}},\ \bibinfo {pages}
  {162--171} (\bibinfo {year} {1951})}\BibitemShut {NoStop}%
\bibitem [{\citenamefont {Bowles}\ and\ \citenamefont
  {Mackenzie}(1954)}]{bowles1954crystallography}%
  \BibitemOpen
  \bibfield  {author} {\bibinfo {author} {\bibfnamefont {J.}~\bibnamefont
  {Bowles}}\ and\ \bibinfo {author} {\bibfnamefont {J.}~\bibnamefont
  {Mackenzie}},\ }\bibfield  {title} {\enquote {\bibinfo {title} {The
  crystallography of martensite transformations i},}\ }\href@noop {} {\bibfield
   {journal} {\bibinfo  {journal} {Acta metallurgica}\ }\textbf {\bibinfo
  {volume} {2}},\ \bibinfo {pages} {129--137} (\bibinfo {year}
  {1954})}\BibitemShut {NoStop}%
\bibitem [{\citenamefont {Mackenzie}\ and\ \citenamefont
  {Bowles}(1954)}]{mackenzie1954crystallography}%
  \BibitemOpen
  \bibfield  {author} {\bibinfo {author} {\bibfnamefont {J.}~\bibnamefont
  {Mackenzie}}\ and\ \bibinfo {author} {\bibfnamefont {J.}~\bibnamefont
  {Bowles}},\ }\bibfield  {title} {\enquote {\bibinfo {title} {The
  crystallography of martensite transformations ii},}\ }\href@noop {}
  {\bibfield  {journal} {\bibinfo  {journal} {Acta Metallurgica}\ }\textbf
  {\bibinfo {volume} {2}},\ \bibinfo {pages} {138--147} (\bibinfo {year}
  {1954})}\BibitemShut {NoStop}%
\bibitem [{\citenamefont {Wechsler}, \citenamefont {Lieberman},\ and\
  \citenamefont {TA}(1953)}]{Wechsler1953}%
  \BibitemOpen
  \bibfield  {author} {\bibinfo {author} {\bibfnamefont {M.}~\bibnamefont
  {Wechsler}}, \bibinfo {author} {\bibfnamefont {D.}~\bibnamefont {Lieberman}},
  \ and\ \bibinfo {author} {\bibfnamefont {R.}~\bibnamefont {TA}},\ }\href@noop
  {} {\bibfield  {journal} {\bibinfo  {journal} {Transactions of the American
  Institue of Mining and Metallurgical Engineers}\ }\textbf {\bibinfo {volume}
  {197}},\ \bibinfo {pages} {1503} (\bibinfo {year} {1953})}\BibitemShut
  {NoStop}%
\bibitem [{\citenamefont {Wechsler}, \citenamefont {TA},\ and\ \citenamefont
  {Lieberman}(1960)}]{Wechsler1960}%
  \BibitemOpen
  \bibfield  {author} {\bibinfo {author} {\bibfnamefont {M.}~\bibnamefont
  {Wechsler}}, \bibinfo {author} {\bibfnamefont {R.}~\bibnamefont {TA}}, \ and\
  \bibinfo {author} {\bibfnamefont {D.}~\bibnamefont {Lieberman}},\ }\href@noop
  {} {\bibfield  {journal} {\bibinfo  {journal} {Transactions of the American
  Institue of Mining and Metallurgical Engineers}\ }\textbf {\bibinfo {volume}
  {218}},\ \bibinfo {pages} {202} (\bibinfo {year} {1960})}\BibitemShut
  {NoStop}%
\bibitem [{\citenamefont {Therrien}\ and\ \citenamefont
  {Stevanović}(2020)}]{therrien_2020}%
  \BibitemOpen
  \bibfield  {author} {\bibinfo {author} {\bibfnamefont {F.}~\bibnamefont
  {Therrien}}\ and\ \bibinfo {author} {\bibfnamefont {V.}~\bibnamefont
  {Stevanović}},\ }\href@noop {} {\enquote {\bibinfo {title} {Unifying
  description of the martensitic phase transformation from the minimization of
  atomic displacements},}\ } (\bibinfo {year} {2020}),\ \Eprint
  {http://arxiv.org/abs/1912.11915} {arXiv:1912.11915 [cond-mat.mtrl-sci]}
  \BibitemShut {NoStop}%
\bibitem [{\citenamefont {Date}\ and\ \citenamefont
  {Nagi}(2016)}]{date2016gpu}%
  \BibitemOpen
  \bibfield  {author} {\bibinfo {author} {\bibfnamefont {K.}~\bibnamefont
  {Date}}\ and\ \bibinfo {author} {\bibfnamefont {R.}~\bibnamefont {Nagi}},\
  }\bibfield  {title} {\enquote {\bibinfo {title} {Gpu-accelerated hungarian
  algorithms for the linear assignment problem},}\ }\href@noop {} {\bibfield
  {journal} {\bibinfo  {journal} {Parallel Computing}\ }\textbf {\bibinfo
  {volume} {57}},\ \bibinfo {pages} {52--72} (\bibinfo {year}
  {2016})}\BibitemShut {NoStop}%
\bibitem [{\citenamefont {Nahor}\ and\ \citenamefont
  {Kaplan}(2016)}]{nahor_2016}%
  \BibitemOpen
  \bibfield  {author} {\bibinfo {author} {\bibfnamefont {H.}~\bibnamefont
  {Nahor}}\ and\ \bibinfo {author} {\bibfnamefont {W.~D.}\ \bibnamefont
  {Kaplan}},\ }\bibfield  {title} {\enquote {\bibinfo {title} {Structure of the
  equilibrated ni(111)-ysz(111) solid–solid interface},}\ }\href {\doibase
  10.1111/jace.14015} {\bibfield  {journal} {\bibinfo  {journal} {Journal of
  the American Ceramic Society}\ }\textbf {\bibinfo {volume} {99}},\ \bibinfo
  {pages} {1064--1070} (\bibinfo {year} {2016})}\BibitemShut {NoStop}%
\bibitem [{\citenamefont {Li}\ \emph {et~al.}(2016)\citenamefont {Li},
  \citenamefont {Chen}, \citenamefont {Zang},\ and\ \citenamefont
  {Feng}}]{Li_2016}%
  \BibitemOpen
  \bibfield  {author} {\bibinfo {author} {\bibfnamefont {L.}~\bibnamefont
  {Li}}, \bibinfo {author} {\bibfnamefont {Z.}~\bibnamefont {Chen}}, \bibinfo
  {author} {\bibfnamefont {Y.}~\bibnamefont {Zang}}, \ and\ \bibinfo {author}
  {\bibfnamefont {S.}~\bibnamefont {Feng}},\ }\bibfield  {title} {\enquote
  {\bibinfo {title} {Atomic-scale characterization of si(110)/6h-sic(0001)
  heterostructure by hrtem},}\ }\href {\doibase
  https://doi.org/10.1016/j.matlet.2015.10.017} {\bibfield  {journal} {\bibinfo
   {journal} {Materials Letters}\ }\textbf {\bibinfo {volume} {163}},\ \bibinfo
  {pages} {47 -- 50} (\bibinfo {year} {2016})}\BibitemShut {NoStop}%
\bibitem [{Note3()}]{Note3}%
  \BibitemOpen
  \bibinfo {note} {Our algorithm does not differentiate between $\protect \text
  {OR}_1$ and $\protect \text {OR}_2$ because they are identical
  in-plane.}\BibitemShut {Stop}%
\bibitem [{\citenamefont {Ding}\ \emph {et~al.}(2006)\citenamefont {Ding},
  \citenamefont {Zhou}, \citenamefont {He},\ and\ \citenamefont
  {Zha}}]{ding2006r}%
  \BibitemOpen
  \bibfield  {author} {\bibinfo {author} {\bibfnamefont {C.}~\bibnamefont
  {Ding}}, \bibinfo {author} {\bibfnamefont {D.}~\bibnamefont {Zhou}}, \bibinfo
  {author} {\bibfnamefont {X.}~\bibnamefont {He}}, \ and\ \bibinfo {author}
  {\bibfnamefont {H.}~\bibnamefont {Zha}},\ }\bibfield  {title} {\enquote
  {\bibinfo {title} {R 1-pca: rotational invariant l 1-norm principal component
  analysis for robust subspace factorization},}\ }in\ \href@noop {} {\emph
  {\bibinfo {booktitle} {Proceedings of the 23rd international conference on
  Machine learning}}}\ (\bibinfo {organization} {ACM},\ \bibinfo {year}
  {2006})\ pp.\ \bibinfo {pages} {281--288}\BibitemShut {NoStop}%
\bibitem [{\citenamefont {Nie}\ \emph {et~al.}(2010)\citenamefont {Nie},
  \citenamefont {Huang}, \citenamefont {Cai},\ and\ \citenamefont
  {Ding}}]{nie2010efficient}%
  \BibitemOpen
  \bibfield  {author} {\bibinfo {author} {\bibfnamefont {F.}~\bibnamefont
  {Nie}}, \bibinfo {author} {\bibfnamefont {H.}~\bibnamefont {Huang}}, \bibinfo
  {author} {\bibfnamefont {X.}~\bibnamefont {Cai}}, \ and\ \bibinfo {author}
  {\bibfnamefont {C.~H.}\ \bibnamefont {Ding}},\ }\bibfield  {title} {\enquote
  {\bibinfo {title} {Efficient and robust feature selection via joint
  $\mathscr{l}2,1$-norms minimization},}\ }in\ \href
  {http://papers.nips.cc/paper/3988-efficient-and-robust-feature-selection-via-joint-l21-norms-minimization.pdf}
  {\emph {\bibinfo {booktitle} {Advances in Neural Information Processing
  Systems 23}}},\ \bibinfo {editor} {edited by\ \bibinfo {editor}
  {\bibfnamefont {J.~D.}\ \bibnamefont {Lafferty}}, \bibinfo {editor}
  {\bibfnamefont {C.~K.~I.}\ \bibnamefont {Williams}}, \bibinfo {editor}
  {\bibfnamefont {J.}~\bibnamefont {Shawe-Taylor}}, \bibinfo {editor}
  {\bibfnamefont {R.~S.}\ \bibnamefont {Zemel}}, \ and\ \bibinfo {editor}
  {\bibfnamefont {A.}~\bibnamefont {Culotta}}}\ (\bibinfo  {publisher} {Curran
  Associates, Inc.},\ \bibinfo {year} {2010})\ pp.\ \bibinfo {pages}
  {1813--1821}\BibitemShut {NoStop}%
\end{thebibliography}
\end{document}